\documentclass[11pt,a4paper,tikz,externalize]{custom_article} 

\newcommand{\RR}{\mathbb{R}}          % Real space
\newcommand{\Euc}{\mathcal{E}}        % Euclidean space
\newcommand{\FEuc}{\mathcal{R}(\Euc)} % Frame bundle
\newcommand{\GL}{\mathrm{GL}}         % General linear group
\newcommand{\SL}{\mathrm{SL}}         % Special linear group
\newcommand{\SO}{\mathrm{SO}}         % Special Orthogonal group
\newcommand{\subG}{\mathcal{G}}       % Matrix sub group
\newcommand{\vol}{\mathrm{vol}}   % Volume form
\newcommand{\dd}{\mathrm{d}}      % Exterior derivative
\newcommand{\Id}{\mathbf{1}}      % Matrix identity

\newcommand{\BB}{\mathcal{B}}           % Body
\newcommand{\bb}{\bm{b}}                % Point of the body
\newcommand{\codef}{\Omega}              % Config Déformée
\newcommand{\codefp}{\tilde{\codef}}      % Config Déformée 2
\newcommand{\coref}{\codef_0}             % Config Référence
\newcommand{\pdef}{p}                   % Plongement déformé
\newcommand{\pref}{\pdef_0}             % Plongement de Référence
\newcommand{\metDef}{\mathbf{q}}        % Metrique classique déformée
\newcommand{\metRef}{\metDef}        % Metrique classique de référence

\newcommand{\Xref}{\bm{X}}              % Reference point
\newcommand{\Xrefp}{\tilde{\Xref}}      % Reference point 2
\newcommand{\Xdef}{\bm{x}}              % Deformed point
\newcommand{\Xdefp}{\tilde{\Xdef}}       % Deformed point 2
\newcommand{\trans}{\varphi}            % Transformation
\newcommand{\gradT}{\mathbf{F}}         % Gradient de la transformation

\newcommand{\CG}{\mathbf{C}}            % Cauchy Green
\newcommand{\EnTot}{\mathcal{W}}      % Free energy 
\newcommand{\EnLoc}{w}             % Free energy density
\newcommand{\masref}{\rho_0}            % Mass density on the reference configuration

\newcommand{\Codef}{\mathcal{K}}        % Generalised deformed configuration
\newcommand{\Codefp}{\tilde{\Codef}}    % Generalised deformed configuration 2
\newcommand{\Coref}{\Codef_0}           % Generalised reference configuration
\newcommand{\Corefp}{\Codefp_0}         % Generalised reference configuration 2 
\newcommand{\Pdef}{P}                   % Generalised deformed placement 
\newcommand{\Pref}{\Pdef_0}             % Generalised reference placement
\newcommand{\Prefp}{\tilde{\Pdef}_0}    % Generalised reference placement 2

\newcommand{\Sdef}{\mathcal{R}}             % Associated moving frame on the deformed configuration
\newcommand{\Sdefp}{\tilde{\Sdef}}          % Associated moving frame on the deformed configuration 2
\newcommand{\Sref}{\Sdef_0}                 % Associated moving frame on the reference configuration
\newcommand{\Srefp}{\tilde{\Sdef}_0}        % Associated moving frame on the reference configuration 2
\newcommand{\Sglob}{\Sdef^{\mathrm{can}}}   % Canonical moving frame
\newcommand{\fpoint}{R}                     % Pointwise frame
\newcommand{\fref}{\fpoint_{\Xref}}           % Point-wise frame on the reference configuration
\newcommand{\frefp}{\tilde{\fpoint}_{\Xref}}  % Point-wise frame on the reference configuration 2
\newcommand{\fdef}{\fpoint_{\Xdef}}           % Point-wise frame on the deformed configuration
\newcommand{\fdefp}{\tilde{\fpoint}_{\Xdef}}  % Point-wise frame on the deformed configuration 2
\newcommand{\gref}{\mathbf{g}_0}              % Matrix coordinate of a frame on the reference configuration
\newcommand{\grefp}{\tilde{\mathbf{g}}_0}          % Matrix coordinate of a frame on the reference configuration 2
\newcommand{\gdefp}{\mathbf{g}}              % Matrix coordinate of a frame on the deformed configuration
\newcommand{\gdefpp}{\tilde{\mathbf{g}}}          % Matrix coordinate of a frame on the deformed configuration 2
\newcommand{\Gref}{\mathbf{G}_0}            % Matrix coordinate of a moving frame on the reference configuration
\newcommand{\Gdef}{\mathbf{G}}            % Matrix coordinate of a moving frame on the deformed configuration
\newcommand{\VV}{\bm{V}}                    % Vecteur repère référence
\newcommand{\vv}{\bm{v}}                    % Vecteur repère déformé 
\newcommand{\vvp}{\tilde{\bm{v}}}           % Vecteur repère déformé 2
\newcommand{\LocFra}{\vv_i}                 % Base repère défomé 
\newcommand{\LocFrap}{\vvp_i}               % Base repère déformé 2

\newcommand{\Trans}{\Phi}                   % Generalised deformation
\newcommand{\Transp}{\tilde{\Trans}}        % Generalised deformation 2
\newcommand{\tframe}{\boldsymbol{\chi}}     % Micro-deformation
\newcommand{\Tframe}{[\tframe]}             % Matrix representation of the micro-deformation
\newcommand{\TransR}{\Trans_{\trans}}       % Convective generalised deformation
\newcommand{\TransId}{\Trans_{\tframeId}}   % Twisting generalised deformation
\newcommand{\tframeId}{\boldsymbol{\kappa}} % Micro-twist
\newcommand{\TframeX}{[\tframe_{\Xref}]}    % Representation matricielle Micro-transformation ponctuelle
\newcommand{\EnTotC}{\EnTot^{\mathrm{C}}}       % Free energy trivialised
\newcommand{\EnLocC}{\EnLoc^{\mathrm{C}}}              % Free energy density trivialised
\newcommand{\matA}{\mathbf{A}}                  % Matrice Change Gauge
\newcommand{\matAL}{\tilde{\mathbf{A}}}         % Local change of gauge on the (reference) configuration
\newcommand{\matALp}{\tilde{\mathbf{A}}'}         % Local change of gauge on the (deformed) configuration
\newcommand{\diff}{\trans}                           % Diffeomorphism of \Euc
\newcommand{\diffI}{f}                           % Diffeomorphism inv of \Euc
\newcommand{\GA}[1]{\mathrm{GA}_{#1}(\FEuc)}    % Gauge group
\newcommand{\Diff}{\Trans}                        % Diffeomorphism of the frame bundle
\newcommand{\DiffA}{\Diff_{\matAL}}             % Diffeomorphism of the frame bundle
\newcommand{\DiffAp}{\Diff_{\matALp}}             % Diffeomorphism of the frame bundle 2
\newcommand{\tdiff}{\tframe}                       % Induced diffeo
\newcommand{\TdiffX}{[\tdiff_{\Xref}]}          % Pointwise Matrix Induced diffeo
\newcommand{\MM}{\mathcal{M}}           % Base manifold
\newcommand{\FMM}{\mathcal{R}(\MM)}     % Frame bundle
\newcommand{\fmm}{R_{m}}                % Frame
\newcommand{\act}{\mathbf{R}}           % Action mult
\newcommand{\Field}{\Psi}
\newcommand{\Fieldp}{\tilde{\Field}}

\newcommand{\funcR}{\phi}

\newcommand{\vtrans}{\bm{b}}            % Translation
\newcommand{\Rtrans}{\mathbf{Q}}        % Rotation
\title[Geometric theory of generalised continua]{Geometric theory of generalised continua\\ using moving frames}

\author{C. Ecker}
\address[Clément Ecker]{Université Paris-Saclay, CentraleSupélec, ENS Paris Saclay, CNRS, LMPS - Laboratoire de Mécanique Paris-Saclay, 91190, Gif-sur-Yvette, France}
\email{clement.ecker@ens-paris-saclay.fr}

\author{B. Kolev}
\address[Boris Kolev]{Université Paris-Saclay, CentraleSupélec, ENS Paris Saclay, CNRS, LMPS - Laboratoire de Mécanique Paris-Saclay, 91190, Gif-sur-Yvette, France}
\email{boris.kolev@ens-paris-saclay.fr}

\date{\today}%

\hypersetup{
    pdftitle={Geometric theory of generalised continua using moving frames},
    pdfauthor={Clément Ecker and Boris Kolev},
    pdfsubject={MSC 2020: 74, 58D, 20E07},
    pdfkeywords={Moving frames, Micromorphic media, Frame bundle, Gauge theory},
    pdflang=eng
}

\newcommand{\insertAbstract}{
    \begin{abstract}
        Generalised continuum theories couple the macroscopic deformation and the micro-/meso-scopic deformation of an underlying micro-structure. They can account for internal length-scale effects and higher-order mechanical loadings absent from classical Cauchy elasticity and has shown its efficiency in modelling metamaterials. This has led to a proliferation of higher-grade and higher-order models (\textit{e.g.} strain-gradient, micromorphic, micro-polar) whose underlying kinematic reduction strategies, in particular to reduce the number of material parameters, are rarely identifiable from the free energy alone. It leaves two open issues: the lack of a systematic criterion for selecting an appropriate model, and a persistent ambiguity regarding the physical status of the local frames ("directors of matter") used in their kinematic description. Are they physical quantities describing the change of state of the given micro-structure or arbitrary kinematic descriptors of its deformation ? This work addresses both questions through a gauge-theoretic formulation of generalised continua in finite strains. While relying on tools and modelling choices consistent with the existing geometric literature on continuum mechanics, the present approach departs from it in its objectives, aiming at a unifying classification of available mechanical models rather than the description of a specific microstructural phenomenon such as defects. In this work, generalised configurations are defined as moving frames over classical configurations, and invariance with respect to the reference generalised configuration is shown to be a necessary and sufficient condition for a gauge invariance, recovering the micromorphic theory as the general-purpose theory of the deformation of arbitrary directors of matter. A systematic classification of first-order generalised media follows from structural group reduction, while strain-gradient continua are recovered through convected frames, and finally constrained media (\textit{e.g.} couple-stress) are addressed.
    \end{abstract}
}

\tikzset{
arrowRep/.style={ultra thick, toolviolet, -{Stealth[scale=1.2]}},
arrete/.style={ultra thick, gray},
pics/repere/.style={
        code={
                \draw[arrowRep] (0,0,0) -- (1,0,0) node[right] {};
                \draw[arrowRep] (0,0,0) -- (0,1,0) node[above] {};
                \draw[arrowRep] (0,0,0) -- (0,0,1.2) node[below left] {};
            }
    },
pics/monCube/.style={
        code={
                \begin{scope}[transform shape]
                    \draw[arrete] (1,0,0) -- (1,1,0) -- (0,1,0);
                    \draw[arrete] (1,1,0) -- (1,1,1) -- (0,1,1) -- (0,1,0);
                    \draw[arrete] (1,0,1) -- (1,1,1);
                    \draw[arrete] (0,0,1) -- (0,1,1);
                    \draw[arrete] (0,0,1) -- (1,0,1) -- (1,0,0);
                    \draw[arrete] (0,0,0) -- (1,0,0);
                    \draw[arrete] (0,0,0) -- (0,1,0);
                    \draw[arrete] (0,0,0) -- (0,0,1);
                \end{scope}
            }
    }
}

\newcommand{\insertDiagConvect}{
    \tikzsetnextfilename{diagConvect}
    \begin{tikzpicture}[>=Stealth, scale=1.3]

        \begin{scope}
            \draw[thick, fill=gray!10, draw=black!70]
            plot[smooth cycle, tension=0.7] coordinates {(-1.5,-1.2) (-1.2,1.2) (1.2,1) (1.5,-0.8) (0,-1.4)};

            \coordinate (X) at (0, 0);
            \draw[thick] (X) to[out=90, in=315] (-0.2,0.6);
            \draw[thick] (X) to[out=270, in=0] (-0.8,-0.6);
            \node[below] at (-0.8,-0.6) {$\Xref(s)$};

            \fill[toolred] (X) circle (2pt) node[below right, xshift=2pt, toolred] {$\Xref$};
            \draw[->, ultra thick, toolblue] (X) -- ++(0,0.9) node[right, toolblue] {$\VV$ (tangent vector)};

            \node[align=center] at (0,-1.9) {\textbf{Reference configuration} $\coref$};
            \coordinate (top) at (1.7, 0);
        \end{scope}

        \begin{scope}[shift={(6,0)}]
            \draw[thick, fill=gray!10, draw=black!70]
            plot[smooth cycle, tension=0.7] coordinates {(-1.1,-1.3) (-1.4,0.9) (0.9,1.4) (1.8,0.1) (1.1,-1.3)};

            \coordinate (x) at (0, 0);
            \draw[thick] (x) to[out=45, in=180] (1.1,0.4);
            \draw[thick] (x) to[out=225, in=0] (-1,-0.7);
            \node[below right] at (-1,-0.7) {$\Xdef(s) = \trans(\Xref(s))$};

            \fill[toolred] (x) circle (2pt) node[below right, xshift=2pt, toolred] {$\Xdef$};
            \draw[->, ultra thick, toolblue] (x) -- ++(0.8,0.8) node[left, toolblue, align=left] {$\vv$ (convected vector)};

            \node[align=center] at (0,-1.9) {\textbf{Deformed configuration} $\codef$};
            \coordinate (topf) at (-1.7, 0);
        \end{scope}

        \draw[->, thick, toolred] (top) to node[above, yshift=2pt, toolred] {$\trans$ (deformation)} (topf);

    \end{tikzpicture}
}

\newcommand{\insertDiagBeams}{
    \tikzsetnextfilename{diagBeams}
    \begin{tikzpicture}[
        scale = 1.3,
        beam/.style        = {line width=1pt, line cap=round},
        latticenode/.style = {circle, fill=toolred, draw=toolred,
                inner sep=0pt, minimum size=4.5pt},
        director/.style    = {-{Latex[length=2.2mm,width=1.4mm]},
        draw=toolviolet, line width=0.9pt}
        ]

        \begin{scope}[xshift=0cm, yshift=0cm]
            \foreach \i in {0,...,2}{
                    \foreach \j in {0,...,2}{
                            \coordinate (N\i\j) at (\i,\j);
                        }
                }
            \foreach \j in {0,...,2}{
                    \draw[beam] (N0\j) -- (N1\j) -- (N2\j);
                }
            \foreach \i in {0,...,2}{
                    \draw[beam] (N\i0) -- (N\i1) -- (N\i2);
                }
            \foreach \i in {0,1}{
                    \foreach \j in {0,1}{
                            \pgfmathtruncatemacro{\ip}{\i+1}
                            \pgfmathtruncatemacro{\jp}{\j+1}
                            \draw[beam] (N\i\j) -- (N\ip\jp);
                            \draw[beam] (N\ip\j) -- (N\i\jp);
                        }
                }
            \foreach \i in {0,...,2}{
                    \foreach \j in {0,...,2}{
                            \node[latticenode] at (N\i\j) {};
                        }
                }
            \node at (1,-0.8) {(a)};
        \end{scope}

        \begin{scope}[xshift=4.5cm, yshift=1cm]
            \def\L{0.9}   % longueur des demi-poutres
            \def\d{0.42}  % longueur des vecteurs directeurs

            \coordinate (C)  at (0,0);
            \coordinate (Nn) at (0, \L);
            \coordinate (Nne) at (\L,\L);
            \coordinate (Ne) at ( \L,0);
            \coordinate (Nse) at (\L,-\L);

            \draw[beam] (C) -- (Nn);
            \draw[beam] (C) -- (Nne);
            \draw[beam] (C) -- (Ne);
            \draw[beam] (C) -- (Nse);

            \foreach \P in {C,Nn,Nne,Ne,Nse}{
                    \node[latticenode] at (\P) {};
                }

            \foreach \P in {C,Nn,Nne,Ne,Nse}{
                    \draw[director] (\P) -- ++(\d,0);
                    \draw[director] (\P) -- ++(0,\d);
                }

            \node at (0,-1.6) {(b)};
        \end{scope}

        \begin{scope}[xshift=9cm, yshift=1cm]
            \def\L{0.9}
            \def\d{0.42}
            \def\rotang{37}   % angle de rotation du noeud central (degrés)

            \coordinate (C)  at (0,0);
            \coordinate (Nn) at (0, \L);
            \coordinate (Nne) at (\L,\L);
            \coordinate (Ne) at ( \L,0);
            \coordinate (Nse) at (\L,-\L);

            \draw[beam] (C) to[out=90+\rotang,  in=-90] (Nn);
            \draw[beam] (C) to[out=45+\rotang, in=225]  (Nne);
            \draw[beam] (C) to[out=\rotang,     in=180] (Ne);
            \draw[beam] (C) to[out=315+\rotang, in=135]   (Nse);

            \foreach \P in {Nn,Nne,Ne,Nse}{
                    \node[latticenode] at (\P) {};
                }
            \foreach \P in {Nn,Nne,Ne,Nse}{
                    \draw[director] (\P) -- ++(\d,0);
                    \draw[director] (\P) -- ++(0,\d);
                }

            \begin{scope}[rotate around={\rotang:(C)}]
                \draw[director] (C) -- ++(\d,0);
                \draw[director] (C) -- ++(0,\d);
            \end{scope}

            \node[latticenode] at (C) {};

            \node at (0,-1.6) {(c)};
        \end{scope}

    \end{tikzpicture}

}

\newcommand{\insertDiagLattice}{
    \tikzsetnextfilename{diagLattice}
    \begin{tikzpicture}[scale=0.6]

        \begin{scope}[shift={(0,0)}]
            \foreach \x in {-2,...,2} {
                    \foreach \y in {-2,...,2} {
                            \fill (\x*1.5, \y*1.5) circle (3pt);
                        }
                }

            \draw[{Stealth[scale=1.2]}-, toolviolet, ultra thick] (1.5, 0) -- (0,0);
            \draw[{Stealth[scale=1.2]}-, toolviolet, ultra thick] (0, 1.5) -- (0,0);

            \node[below, toolviolet] at (0,0) {$\fref$};

            \node[anchor=north, font=\sffamily\Large] at (0, -3.8) {(a) Square lattice};
        \end{scope}

        \begin{scope}[shift={(9,0)}]

            \foreach \i in {-2,...,2} {
                    \foreach \j in {-2,...,2} {
                            \pgfmathtruncatemacro{\val}{abs(\i+\j)}
                            \ifnum\val<3
                                \fill (\i*1.5 + \j*0.75, \j*1.3) circle (3pt);
                            \fi
                        }}

            \draw[{Stealth[scale=1.2]}-, toolviolet, ultra thick] (1.5, 0) -- (0,0);
            \draw[{Stealth[scale=1.2]}-, toolviolet, ultra thick] (0.75, 1.3) -- (0,0);

            \node[below, toolviolet] at (0,0) {$\frefp$};

            \node[anchor=north, font=\sffamily\Large] at (0.3, -3.8) {(b) Hexagonal lattice};
        \end{scope}

    \end{tikzpicture}
}

\newcommand{\insertDiagScale}{
    \tikzsetnextfilename{diagScales}
    \definecolor{darkgray}{RGB}{110,113,115}
    \definecolor{scaleblue}{RGB}{0,174,239}
    \definecolor{scaleorange}{RGB}{241,90,41}
    \begin{tikzpicture}[every node/.style={inner sep=0, outer sep=0}, scale=0.8]

        \def\valRad{0.2}
        \def\valRatioCell{1.4}
        \def\nbxRow{12}

        \def\celllength{
            2*\valRatioCell*\valRad
        }
        \def\boxlength{
            \celllength*0.866*\nbxRow
        }
        \begin{scope}
            \clip (0,0) rectangle (\boxlength,\boxlength);
            \fill[darkgray] (0,0) rectangle (\boxlength,\boxlength);

            \foreach \row in {-1,...,12} {
                    \foreach \col in {-1,...,12} {
                            \pgfmathsetmacro{\xshift}{mod(\row,2)*\celllength/2}
                            \fill[white] (\col*\celllength + \xshift, \row*\celllength*0.866) circle (\valRad);
                        }
                }
        \end{scope}
        \draw[thick] (0,0) rectangle (\boxlength,\boxlength);
        \node[anchor=north, font=\sffamily\bfseries] at (3,-0.2) {Macroscale};

        \coordinate (macroCenter) at (7*\celllength, 6*\celllength*0.866);
        \def\macroRadiusRatio{2}
        \def\macroRadius{\macroRadiusRatio*\celllength/2}

        \draw[scaleblue, ultra thick] (macroCenter) circle (\macroRadius);

        \coordinate (mesoCenter) at (10, 3.2);
        \def\mesoCellRadius{3*\valRad}
        \def\mesoRadius{\macroRadiusRatio*\valRatioCell*\mesoCellRadius}
        \def\celllengthMeso{2*\mesoCellRadius*\valRatioCell}

        \draw[scaleblue, thick] ($(macroCenter)+(45:\macroRadius)$) -- ($(mesoCenter)+(135:\mesoRadius)$);
        \draw[scaleblue, thick] ($(macroCenter)+(-45:\macroRadius)$) -- ($(mesoCenter)+(-135:\mesoRadius)$);

        \begin{scope}
            \clip (mesoCenter) circle (\mesoRadius);
            \fill[darkgray] ($(mesoCenter)-(\mesoRadius,\mesoRadius)$) rectangle ($(mesoCenter)+(\mesoRadius,\mesoRadius)$);

            \foreach \row in {-1,...,1} {
                    \foreach \col in {-1,...,1} {
                            \pgfmathsetmacro{\xshift}{mod(\row,2)*\celllengthMeso/2}
                            \fill[white] ($(mesoCenter) + (\col*\celllengthMeso + \xshift, \row*\celllengthMeso*0.866)$) circle (\mesoCellRadius);
                        }
                }
        \end{scope}
        \draw[scaleblue, line width=3pt] (mesoCenter) circle (\mesoRadius);
        \node[anchor=north, font=\sffamily\bfseries] at (10, 1) {Mesoscale};

        \coordinate (mesoSelect) at ($(mesoCenter)+(14.8:0.95)$);
        \def\mesoSelectRadius{0.18}

        \draw[scaleorange, thick] (mesoSelect) circle (\mesoSelectRadius);

        \coordinate (microCenter) at (14.5, 3.2);
        \def\microRadius{1.0}

        \draw[scaleorange, thick] ($(mesoSelect)+(60:\mesoSelectRadius)$) -- ($(microCenter)+(130:\microRadius)$);
        \draw[scaleorange, thick] ($(mesoSelect)+(-60:\mesoSelectRadius)$) -- ($(microCenter)+(-130:\microRadius)$);

        \begin{scope}
            \clip (microCenter) circle (\microRadius);
            \fill[darkgray] ($(microCenter)-(\microRadius,\microRadius)$) rectangle ($(microCenter)+(\microRadius,\microRadius)$);
        \end{scope}
        \draw[scaleorange, line width=3pt] (microCenter) circle (\microRadius);
        \node[anchor=north, font=\sffamily\bfseries] at (14.5, 1) {Microscale};

    \end{tikzpicture}
}

\newcommand{\insertDiagCube}{
    \begin{tikzpicture}[
        node distance=0.6cm and 0.3cm,
        label node/.style={align=center, font=\small},
        arrow/.style={-{Stealth[scale=1.2]}, thick},
        ptB/.style={circle, fill=toolred, inner sep=2pt},
        labelP/.style={font=\normalsize, text=toolred, anchor=north}
        ]
        \node[ptB] (K0) {};
        \node[ptB, right=5cm of K0] (K) {};

        \node[left=of K0, toolviolet] {$\Coref$};
        \node[right=0.8cm of K, toolviolet] {$\Codef$};
        \node[above=0.8cm of K0, label node] {Reference generalised\\configuration};
        \node[above=0.8cm of K, label node] {Deformed generalised\\configuration};

        \pic[] at ([shift={(-0.5,-0.5,-0.5)}]K0) {monCube};
        \pic[rotate=20, xscale=1.3, yscale=0.9] at ([shift={(-0.5,-0.5,-0.5)}]K) {monCube};

        \draw[arrow] ($(K0) + (1.2,0)$) -- ($(K) + (-0.6,0)$) node[midway, above] {$\Trans$};

        \pic[] at (K0) {repere};
        \pic[rotate=20, xscale=1.3, yscale=0.9] at (K) {repere};

        \node[labelP] at ([yshift=-0.1]K0.south east) {$\Xref$};
        \node[labelP] at ([yshift=-0.1]K.south east) {$\Xdef$};
    \end{tikzpicture}
}

\newcommand{\insertDiagExtension}{
    \tikzsetnextfilename{diagDiagExtension}
    \begin{tikzpicture}[
        node distance=0.4cm and 0.3cm,
        label node/.style={align=center, font=\small},
        arrow/.style={-{Stealth[scale=1.2]}, thick}
        ]
        \node (K0) {};
        \node[right=5cm of K0] (K) {};
        \node[below=3cmof K0] (Kt0) {};
        \node[right=5cm of Kt0] (Kt) {};

        \node[left=of K0, toolviolet] {$\fref$};
        \node[left=of Kt0, toolviolet] {$\frefp=\fref\matA$};
        \node[right=0.8cm of K, toolviolet] {$\fdef$};
        \node[right=0.8cm of Kt, toolviolet] {$\fdefp=\fdef\matA$};
        \node[above=1.2cm of K0, label node] {Reference frame (1)};
        \node[above=1.2cm of K, label node] {Deformed frame (1)};
        \node[below=of Kt0, label node] {Reference frame (2)};
        \node[below=of Kt, label node] {Deformed frame (2)};

        \pic[] at (K0) {monCube};
        \pic[] at (Kt0) {monCube};
        \pic[rotate=20, xscale=1.3, yscale=0.9] at (K) {monCube};
        \pic[rotate=20, xscale=1.3, yscale=0.9] at (Kt) {monCube};

        \draw[arrow] ($(K0) + (1.2,0)$) -- ($(K) + (-0.6,0)$) node[midway, above, toolviolet] {$\Trans$};
        \draw[arrow] ($(Kt0) + (1.2,0)$) -- ($(Kt) + (-0.6,0)$) node[midway, below, toolviolet] {$\Trans$};
        \draw[arrow] ($(K0) + (0.3,-0.5)$) -- ($(Kt0) + (0.3,1.1)$) node[midway, left, toolblue] {$\act_{\matA}$};
        \draw[arrow] ($(K) + (0.3,-0.5)$) -- ($(Kt) + (0.3,1.1)$) node[midway, right, toolblue] {$\act_{\matA}$};

        \pic[] at ([shift={(0.5,0.5,0.5)}]K0) {repere};
        \pic[rotate=30] at ([shift={(0.5,0.5,0.5)}]Kt0) {repere};
        \pic[rotate=20, xscale=1.3, yscale=0.9] at ([shift={(0.5,0.5,0.5)}]K) {repere};
        \pic[rotate=50, xscale=1.3, yscale=0.9] at ([shift={(0.5,0.5,0.5)}]Kt) {repere};
    \end{tikzpicture}
}

\newcommand{\insertDiagGaugeInv}{
    \begin{tikzpicture}[
        node distance=0.4cm and 0.3cm,
        label node/.style={align=center, font=\small},
        arrow/.style={-{Stealth[scale=1.2]}, thick}
        ]
        \node (K0) {};
        \node[right=5cm of K0] (K) {};
        \node[below=3.5cmof K0] (Kt0) {};
        \node[right=5cm of Kt0] (Kt) {};

        \node[left=of K0, toolviolet] {$\Coref$};
        \node[left=of Kt0, toolviolet] {$\Corefp$};
        \node[right=0.8cm of K, toolviolet] {$\Codef$};
        \node[right=0.8cm of Kt, toolviolet] {$\Codefp$};
        \node[above=1.2cm of K0, label node] {Choice of reference\\generalised configuration (1)};
        \node[above=1.2cm of K, label node] {Deformed generalised\\configuration (1)};
        \node[below=of Kt0, label node] {Choice of reference\\generalised configuration (2)};
        \node[below=of Kt, label node] {Deformed generalised\\configuration (2)};

        \pic[] at (K0) {monCube};
        \pic[] at (Kt0) {monCube};
        \pic[rotate=20, xscale=1.3, yscale=0.9] at (K) {monCube};
        \pic[rotate=20, xscale=1.3, yscale=0.9] at (Kt) {monCube};

        \draw[arrow] ($(K0) + (1.2,0)$) -- ($(K) + (-0.6,0)$) node[midway, above] {$\Field$};
        \draw[arrow] ($(Kt0) + (1.2,0)$) -- ($(Kt) + (-0.6,0)$) node[midway, above] {$\DiffA\star\Field$};
        \draw[arrow] ($(K0) + (0.3,-0.5)$) -- ($(Kt0) + (0.3,1.1)$) node[midway, left] {$\DiffA$};
        \draw[arrow] ($(K) + (0.3,-0.5)$) -- ($(Kt) + (0.3,1.1)$) node[midway, right] {$\approx\DiffA$};

        \pic[] at ([shift={(0.5,0.5,0.5)}]K0) {repere};
        \pic[rotate=30] at ([shift={(0.5,0.5,0.5)}]Kt0) {repere};
        \pic[rotate=20, xscale=1.3, yscale=0.9] at ([shift={(0.5,0.5,0.5)}]K) {repere};
        \pic[rotate=50, xscale=1.3, yscale=0.9] at ([shift={(0.5,0.5,0.5)}]Kt) {repere};
    \end{tikzpicture}
}

\newcommand{\insertDiagTangent}{
    \begin{tikzpicture}[
            node distance=2.5cm and 3cm,
            vec/.style={->, >=stealth', ultra thick},
            ptB/.style={circle, fill=toolred, inner sep=2pt},
            vecP/.style={->, >=stealth', ultra thick, toolblue},
            labelV/.style={font=\normalsize, text=toolblue}
        ]

        \node[label] (z) at (0,2.5) {};
        \node[anchor=south, font=\Large] at ([shift={(-0.8,0.6)}]z) {$\RR^3$};

        \draw[vec] (z.west) -- ++(1,0);
        \node[font=\normalsize, anchor=west] at ([shift={(1,0)}]z) {$\bm{e}_1$};
        \draw[vec] (z.south) -- ++(0,1);
        \node[font=\normalsize, anchor=south] at ([shift={(0,1)}]z) {$\bm{e}_2$};
        \draw[vec] (z.north east) -- ++(-0.5,-0.5);
        \node[font=\normalsize, anchor=north] at ([shift={(-0.5,-0.5)}]z) {$\bm{e}_3$};

        \node[ptB, below left=of z] (X) {};
        \node[font=\normalsize, anchor=north, text=toolred] at ([yshift=-0.1]X.south) {$\Xref$};

        \draw[vecP] (X) -- ++(1.5,-0.3);
        \node[labelV, anchor=west] at ([shift={(1.5,-0.3)}]X) {$\VV_1$};
        \draw[vecP] (X) -- ++(0.5,0.5);
        \node[labelV, anchor=west] at ([shift={(0.5,0.5)}]X) {$\VV_2$};
        \draw[vecP] (X) -- ++(-0.5,1.3);
        \node[labelV, anchor=south] at ([shift={(-0.5,1.3)}]X) {$\VV_3$};

        \node[ptB, below right=of z] (x) {};
        \node[font=\normalsize, anchor=north, text=toolred, anchor=west] at ($(x) + (-0.25,-0.55)$) {$\Xdef=\diff(\Xref)$};

        \draw[vecP] (x) -- ++(1.5,-0.3);
        \node[labelV, anchor=west] at ([shift={(1.5,-0.3)}]x) {$\vv_1$};
        \draw[vecP] (x) -- ++(0.5,0.5);
        \node[labelV, anchor=west] at ([shift={(0.5,0.5)}]x) {$\vv_2$};
        \draw[vecP] (x) -- ++(-0.5,1.3);
        \node[labelV, anchor=south] at ([shift={(-0.5,1.3)}]x) {$\vv_3$};

        \draw[vec, toolviolet] ($(z) + (-0.2,0.3)$) to [out=180, in=90]
        node[font=\large, text=toolviolet, midway, above, yshift=2pt] {$\fref$} ([shift={(0,0.5)}]X);
        \draw[vec, toolviolet] ($(z) + (0.2,0.3)$) to [out=0, in=90]
        node[font=\large, text=toolviolet, midway, above, yshift=2pt, anchor=west] {$\fdef=\Diff(\fref)$}  ([shift={(0,0.5)}]x);
        \draw[vec, toolviolet] ($(X) + (0.3,-0.4)$) to [out=-30, in=210]
        node[font=\large, text=toolviolet, midway, above, yshift=2pt] {$\Diff$}  ([shift={(-0.3,-0.4)}]x);

        \draw[vec, toolred] (z) -- ++(0.8,0.8);
        \node[toolred, anchor=west] at ([shift={(1,1)}]z) {$\bm{e}$};
        \draw[vec, toolred] (X) -- ++(1,0.1);
        \node[toolred, anchor=west] (dX) at ([shift={(1,0.1)}]X) {$\delta\Xref$};
        \draw[vec, toolred] (x) -- ++(1,0.1);
        \node[toolred, anchor=west] at ([shift={(1,0.1)}]x) {$\delta\Xdef$};
        \draw[vec, toolred] ($(dX.north) + (0,0)$) to [out=10, in=170]
        node[font=\large, text=toolred, midway, above, yshift=2pt] {$\tdiff_{\Xref}$}  ([shift={(-0.2,0.2)}]x);

    \end{tikzpicture}
}

\newcommand{\insertDiagObj}{
    \begin{tikzpicture}[
        node distance=2.5cm and 3cm,
        cloud/.style={draw, ellipse, minimum width=2.5cm, minimum height=1.5cm, fill=blue!5},
        cloudd/.style={draw, ellipse, minimum width=3.0cm, minimum height=1.0cm, fill=blue!5},
        arrow/.style={-{Stealth[scale=1.2]}, thick},
        label node/.style={align=center, font=\small}
        ]

        \node[cloud] (O0) {$\coref$};
        \node[label node, above=0.2cm of O0] {Reference configuration};

        \node[cloudd, right=of O0] (O) {$\codef$};
        \node[label node, above=0.2cm of O] {Deformed configuration\\seen by observer (1)};

        \node[cloudd, below=of O, rotate=20] (Ot) {$\codefp$};
        \node[label node, below=0.2cm of Ot] {Deformed configuration\\seen by observer (2)};

        \draw[arrow] (O0) -- (O) node[midway, above] {$\trans$};
        \draw[arrow] (O0) -- (Ot) node[midway, below] {$\diffI_*\trans$};

        \draw[arrow] (O) -- (Ot) node[midway, right] {$\diffI$};

    \end{tikzpicture}
}

\newcommand{\insertDiagConfigEx}{
    \tikzsetnextfilename{DiagConfigEx}
    \begin{tikzpicture}[
        node distance=0.5cm and 1.3cm ,
        label/.style={align=center, font=\large},
        point/.style={circle, fill=toolred, inner sep=2pt},
        labelP/.style={font=\normalsize, toolred, anchor=north},
        labelR/.style={font=\normalsize, toolviolet, anchor=south west},
        separator/.style={ultra thick},
        curve/.style={thick},
        emb/.style={thick, -{Stealth[scale=1.2]}},
        labelEmb/.style={font=\large,midway}
        ]

        \node[label] (body) {Body $\BB$ (labels)};
        \node[label, left=of body] (point) {Point};
        \node[label, right=of body] (cur) {Curve};

        \node[point, below=of point] (ptP) {};
        \node[point, below=of cur] (ptC) {};
        \node[point, below=3cm of ptP] (ptPE) {};
        \node[point, below=3cm of ptC] (ptCE) {};

        \node[labelP] at (ptP.south) {$b$};
        \node[labelP] at (ptC.south) {$\bb$};
        \node[labelP] at (ptPE.south) {$\Xdef$};
        \node[labelP] at (ptCE.south) {$\Xdef$};

        \draw[separator] (body.south) -- ($(body.south) + (0, -5)$);
        \draw[separator] ($(ptP) + (-3,-0.8)$) -- ($(ptC) + (3, -0.8)$);

        \pic[rotate=-12, xscale=0.7, yscale=1.1] at (ptPE) {repere};
        \node[labelR] (RPE) at (ptPE.north east) {$\fdef$};
        \pic[yscale=1.5, xscale=0.7] at (ptCE) {repere};
        \node[labelR] (RCE) at (ptCE.north east) {$\fdef$};

        \def\leng{1.3}
        \draw[curve] ($(ptC) + (-\leng,0)$) -- ($(ptC) + (\leng,0)$);
        \draw[curve] ($(ptC) + (-\leng,0.1)$) -- ($(ptC) + (-\leng,-0.1)$);
        \node[font=\normalsize, anchor=north] at ($(ptC) + (-\leng,-0.1)$) {$0$};
        \draw[curve] ($(ptC) + (\leng,0.1)$) -- ($(ptC) + (\leng,-0.1)$);
        \node[font=\normalsize, anchor=north] at ($(ptC) + (\leng,-0.1)$) {$L$};
        \draw[curve] (ptCE) to [out=0, in=-90] ($(ptCE) + (0.9,1.4)$);
        \draw[curve] (ptCE) to [out=180, in=40] ($(ptCE) + (-1.4,-0.5)$);

        \draw[emb, toolred] (ptP) to [out=220, in=110] node[labelEmb,left] {$\pdef$} (ptPE);
        \draw[emb, toolviolet] (ptP) to [out=320, in=80] node[labelEmb,right] {$\Pdef$} (RPE);
        \draw[emb, toolred] (ptC) to [out=220, in=110] node[labelEmb,left] {$\pdef$} (ptCE);
        \draw[emb, toolviolet] (ptC) to [out=320, in=80] node[labelEmb,right] {$\Pdef$} (RCE);

    \end{tikzpicture}
}

\newcommand{\insertDiagConfig}{
    \begin{tikzpicture}[
            >=Stealth,
            scale=1.2,
            fiber/.style={thick},
            frame_vec/.style={->, toolviolet, thick, shorten >=0.1pt},
            label_node/.style={font=\small}
        ]

        \node[label_node] at (2.5, 4) {Body $\BB$ (labels)};
        \draw[fiber] (1, 3.5) -- (4, 3.5);
        \foreach \x/\txt in {1/0, 2.5/\bb, 4/L} {
                \draw[thick] (\x, 3.4) -- (\x, 3.6);
                \node[below] at (\x, 3.4) {$\txt$};
            }
        \fill[toolred] (2.5, 3.5) circle (1.5pt) node[below left, toolred] {};

        \draw[ultra thick] (-1.5, 2.8) -- (7, 2.8);
        \node[font=\huge] (ref) at (0.5, 2.5) {Reference};
        \node[font=\huge] (def) at (5, 2.5) {Deformed};

        \node[toolviolet, label_node, anchor=north] at (ref.south) {$\Coref$ (frames along $\coref$)};
        \node[toolviolet, label_node, anchor=north] at (def.south) {$\Codef$ (frames along $\codef$)};

        \begin{scope}[shift={(-1,0)}]

            \draw[fiber] plot [smooth, tension=1.1] coordinates {
                    (0,1) (0.75, 0.95) (1.5,0.2) (2.25,-0.8) (3,-0.5)
                };
            \node[label_node] at (1.5, -1.2) {$\coref$ (classical configuration)};

            \foreach \x/\y [count=\i] in {0/1, 1.5/0.2, 3/-0.5} {
                    \begin{scope}[shift={(\x,\y)}, rotate={-20*(\i-1)}]
                        \draw[frame_vec] (0,0) -- (0.6,0);
                        \draw[frame_vec] (0,0) -- (0,0.6);
                        \draw[frame_vec] (0,0) -- (-0.3,-0.3);
                        \ifnum\i=2
                            \node[] (xref) at (0,0) {};
                            \fill[toolred] (xref) circle (1.5pt) node[left, xshift=-2pt, yshift=3pt] {$\Xref$};
                            \node[toolviolet, right] at (0.1, 0.3) {$\fref$};
                        \fi
                    \end{scope}
                }
        \end{scope}

        \begin{scope}[shift={(4,0)}]

            \draw[fiber] plot [smooth, tension=1.1] coordinates {
                    (0,0.5) (0.6, -0.1) (1.5,0.2) (2.25,0.4) (2.55,0.8)
                };
            \node[label_node] at (1.5, -1.2) {$\codef$ (classical configuration)};

            \foreach \x/\y [count=\i] in {0/0.5, 1.5/0.2, 2.55/0.8} {
                    \begin{scope}[shift={(\x,\y)}, rotate={-20*(\i-2)}]
                        \draw[frame_vec] (0,0) -- (0.6,0);
                        \draw[frame_vec] (0,0) -- (0,0.6);
                        \draw[frame_vec] (0,0) -- (-0.3,-0.3);
                        \ifnum\i=2
                            \node[] (xdef) at (0,0) {};
                            \fill[toolred] (xdef) circle (1.5pt) node[below] {$\Xdef$};
                            \node[toolviolet, right] (fdef) at (0.1, 0.3) {$\fdef$};
                        \fi
                    \end{scope}
                }
        \end{scope}

        \draw[->, toolred, bend right=30] (xref) to node[below right, align=center, xshift=-15pt] {$\trans$ (deformation)} (xdef);

        \draw[->, toolviolet, thick, bend left=20] (1.5, 0.7) to node[above, xshift=-15pt,align=center] {$\Trans$ \scriptsize $\begin{pmatrix} \text{generalised} \\ \text{transformation} \end{pmatrix}$} (fdef);

        \begin{scope}[shift={(0.5, -2.5)}]
            \draw[thick, ->] (0,0) -- (0,0.7);
            \draw[thick, ->] (0,0) -- (0.7,0);
            \draw[thick, ->] (0,0) -- (-0.3,-0.3);
            \node[right] at (0.2, 0.4) {$\Sglob$ (canonical frame)};
            \node[scale=1.2] at (1.5, -0.4) {$\Euc$ (Euclidean space)};
        \end{scope}

    \end{tikzpicture}
}

\newcommand{\insertDiagConfigMic}{
    \tikzsetnextfilename{DiagConfigMic}
    \begin{tikzpicture}[
            >=Stealth,
            scale=1.2,
            fiber/.style={thick},
            frame_vec/.style={->, toolviolet, thick, shorten >=0.1pt},
            label_node/.style={font=\small}
        ]

        \node[label_node] at (2.5, 4) {Body $\BB$ (labels)};
        \draw[fiber] (1, 3.5) -- (4, 3.5);
        \foreach \x/\txt in {1/0, 2.5/b, 4/L} {
                \draw[thick] (\x, 3.4) -- (\x, 3.6);
                \node[below] at (\x, 3.4) {$\txt$};
            }
        \fill[toolred] (2.5, 3.5) circle (1.5pt) node[below left, toolred] {};

        \draw[ultra thick] (-1.5, 2.8) -- (7, 2.8);
        \node[font=\large] (ref) at (0.5, 2.5) {Reference};
        \node[font=\large] (def) at (5, 2.5) {Deformed};

        \node[toolviolet, label_node, anchor=north] at (ref.south) {$\Coref$ (frames along $\coref$)};
        \node[toolviolet, label_node, anchor=north] at (def.south) {$\Codef$ (frames along $\codef$)};

        \begin{scope}[shift={(-1,0)}]

            \draw[fiber] plot [smooth, tension=1.1] coordinates {
                    (0,1) (0.75, 0.95) (1.5,0.2) (2.25,-0.8) (3,-0.5)
                };
            \node[label_node, toolred] at (1.5, -1.2) {$\coref$ (cla. config.)};

            \foreach \x/\y [count=\i] in {0/1, 1.5/0.2, 3/-0.5} {
                    \begin{scope}[shift={(\x,\y)}, rotate={-20*(\i-1)}]
                        \ifnum\i=2
                            \node[] (xref) at (0,0) {};
                            \fill[toolred] (xref) circle (1.5pt) node[below left, xshift=3pt] {$\Xref$};
                            \node[above, toolblue] at (0.2, 0.6) {$\delta\Xref$};
                            \draw[frame_vec, toolblue] (0,0) -- (0,0.6);
                        \else
                            \draw[frame_vec] (0,0) -- (0,0.6);
                            \draw[frame_vec] (0,0) -- (0.6,0);
                            \draw[frame_vec] (0,0) -- (-0.3,-0.3);
                        \fi
                    \end{scope}
                }
        \end{scope}

        \begin{scope}[shift={(4,0)}]

            \draw[fiber] plot [smooth, tension=1.1] coordinates {
                    (0,0.5) (0.6, -0.1) (1.5,0.2) (2.25,0.4) (2.55,0.8)
                };
            \node[label_node, toolred] at (1.5, -1.2) {$\codef$ (cla. config.)};

            \foreach \x/\y [count=\i] in {0/0.5, 1.5/0.2, 2.55/0.8} {
                    \begin{scope}[shift={(\x,\y)}, rotate={-20*(\i-2)}]
                        \ifnum\i=2
                            \node[] (xdef) at (0,0) {};
                            \fill[toolred] (xdef) circle (1.5pt) node[below right, xshift=-3pt] {$\Xdef$};
                            \node[above, toolblue] (fdef) at (0, 0.53) {$\delta\Xdef$};
                            \draw[frame_vec, toolblue] (0,0) -- (0,0.6);
                        \else
                            \draw[frame_vec] (0,0) -- (0,0.6);
                            \draw[frame_vec] (0,0) -- (0.6,0);
                            \draw[frame_vec] (0,0) -- (-0.3,-0.3);
                        \fi
                    \end{scope}
                }
        \end{scope}

        \draw[->, toolred, bend right=15] (xref) to node[below right, align=center, xshift=-15pt] {$\trans$ (deformation)} (xdef);

        \draw[->, toolblue, thick, bend left=10] (xref) to node[above, xshift=-7pt,align=center, toolblue] {$\tframe_{\Xref}$\\\scriptsize (micro-formation)} (xdef);

        \begin{scope}[shift={(0.5, -2.5)}]
            \draw[thick, ->] (0,0) -- (0,0.7);
            \draw[thick, ->] (0,0) -- (0.7,0);
            \draw[thick, ->] (0,0) -- (-0.3,-0.3);
            \node[right] at (0.2, 0.4) {$\Sglob$ (canonical frame)};
        \end{scope}

    \end{tikzpicture}
}

\newcommand{\insertDiagFrame}{
    \begin{tikzpicture}[scale=1.1,
            vec/.style={->, >=stealth', ultra thick},
            ptB/.style={circle, fill=toolred, inner sep=2pt},
            vecP/.style={->, >=stealth', ultra thick, toolblue},
            labelV/.style={font=\normalsize, text=toolblue}
        ]

        \node[label] (z) at (-2.2,0) {};
        \node[anchor=south, font=\Large] at (-2.2,2) {$\RR^3$};

        \draw[vec] (z.west) -- ++(1,0);
        \node[font=\normalsize, anchor=west] at ([shift={(1,0)}]z) {$\bm{e}_1$};
        \draw[vec] (z.south) -- ++(0,1);
        \node[font=\normalsize, anchor=south] at ([shift={(0,1)}]z) {$\bm{e}_2$};
        \draw[vec] (z.north east) -- ++(-0.5,-0.5);
        \node[font=\normalsize, anchor=north] at ([shift={(-0.5,-0.5)}]z) {$\bm{e}_3$};

        \node[ptB] (x) at (2.2,0) {};
        \node[font=\normalsize, anchor=north, text=toolred] at ([yshift=-0.1]x.south) {$\Xdef$};
        \node[anchor=south, font=\Large] at (2.2,2) {$\Euc$};

        \draw[vecP] (x) -- ++(1.5,-0.3);
        \node[labelV, anchor=west] at ([shift={(1.5,-0.3)}]x) {$\vv_1$};
        \draw[vecP] (x) -- ++(0.5,0.5);
        \node[labelV, anchor=west] at ([shift={(0.5,0.5)}]x) {$\vv_2$};
        \draw[vecP] (x) -- ++(-0.5,1.3);
        \node[labelV, anchor=south] at ([shift={(-0.5,1.3)}]x) {$\vv_3$};

        \draw[vec, toolviolet] (-1.7,0.5) to [out=45, in=135] ([shift={(-0.5,0.5)}]x);
        \node[font=\large, text=toolviolet] at (0,1.4) {$\fdef$};

    \end{tikzpicture}
}            % Tikz diagram
\begin{document}

% -------------------- STRUCTURE --------------------

\insertAbstract
\maketitle
\tableofcontents

% ----------------------------------------------------------------
\section{Introduction}
\label{sec:introduction}
% ----------------------------------------------------------------

\begin{figure}[h]
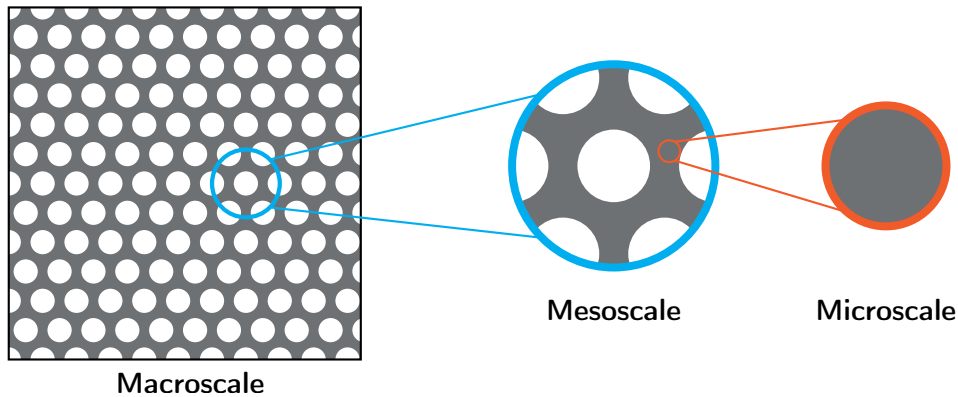

    \centering
    \insertDiagScale
    \caption{Denomination of scales : macroscopic, mesoscopic and microscopic (adapted from~\parencite{PSM+2018}).}
    \label{fig:schScale}
\end{figure}

The classical Cauchy continuum theory, while foundational, possesses intrinsic limitations that prevent it to accurately describes modern complex materials~\parencite{PSM+2018, IS2023, RAC2024}. It cannot account for internal length-scale effects~\parencite{DFI2006, AA2011}, nor can it accommodate higher-order mechanical actions such as body couples or double-forces~\parencite{Ger1973, JIS2013, IS2023}. For example, in the context of wave propagation, classical continuum mechanics fails to capture the dispersive nature and band structures observed in many heterogeneous media~\parencite{BES+2013, MNA+2017}. To address these shortcomings, generalised media theories introduce a coupling between the macroscopic scale and an underlying microscopic (or mesoscopic, see \cref{fig:schScale}) scale~\parencite{PSM+2018, RAC2024}. This need is particularly acute in the study of architected materials (or metamaterials) where the microstructure size is not negligible compared to the macroscopic sample~\parencite{DFI2006, MNA+2017, BEIH2020}.

While the homogenisation of such materials aims to provide a continuous description through rigorous micro-macro identification~\parencite{AMI2017, JS2020, FPF2024}, the success of these procedures relies heavily on the choice of a sufficiently rich macroscopic kinematic target model. The literature offers an abundance of generalised models, which can be broadly categorised as the following~\parencite{Ger1973,FS2006}: higher-grade media such as strain-gradient theories~\parencite{Tou1962, ME1968, Koi1963}, where the free energy density depends on higher gradients of deformation; and higher-order media such as micromorphic, micro-polar, or micro-dilatation theories~\parencite{ES1964a, Min1964, Cow1984, Eri1999}, where additional degrees of freedom are introduced.

The proliferation of theories of generalised continua has led to significant technical and conceptual challenges, particularly in the regime of finite strains, a topic only recently emerging outside theoretical modeling~\parencite{BEIH2020,LFD2024,McA2025}. For practical applications, selecting the appropriate generalised framework a priori for a given physical problem presents a major hurdle~\parencite{Com2023,LSS2026}. One explanation for the large number of models available is a consequence of the rapid explosion of constitutive parameters required by these theories (\textit{e.g.} a fully anisotropic (triclinic) linear strain gradient medium requires $171$ material constants~\parencite{ALH2013}). To render generalised media operational, modelers employ reduction procedures to limit either the number of degrees of freedom or the number of constitutive parameters. Outside the scope of material symmetries, we roughly categorise these simplification pathways into four distinct strategies:
\begin{itemize}
    \item \emph{Microscopic Kinematic Reduction}: Specific degrees of freedom of the micro-volumes are removed. An example is the Cosserat (micro-polar) medium, where all internal non-rigid strain degrees of freedom are forbidden, restricting the microstructure to pure rotations~\parencite{CC1909}.
    \item \emph{Coupled Kinematic Reduction}: Certain microstructural degrees of freedom lose their independence and become entirely determined by the macroscopic deformation field. For instance the strain-gradient model in which the micro-deformation is forced to equal the macroscopic deformation gradient $\gradT$~\parencite{FS2025}.
    \item \emph{Energetic Reduction}: Specific kinematic mechanisms or strain measures are postulated to yield no pointwise energetic cost. In the relaxed micromorphic model, only the $\text{curl}$ of the micro-deformation (\textit{i.e.} only a part of the gradient) contributes to the free energy density~\parencite{NGM+2014}.
    \item \emph{Constitutive Modeling Reduction}: Algebraic relationships are enforced between material parameters, reducing the number of independent parameters. In the micro-foam model, several material parameters of a linear micromorphic medium are linked to a single microstructural length-scale parameter~\parencite{DFI2006}.
\end{itemize}
While the physical and intuitive distinctions between these four reduction modes are transparent when derived from a known microstructure, they become ambiguous within a generic model that can be applied to almost arbitrary micro-structure, which is the goal of a general purpose continuum theory~\parencite{Ger1973}. A path toward a classification of the kinematics of generalised continua is the main applicative objective of this paper.

Modern computational mechanics often relies on axiomatic approaches where the media explicit kinematic is bypassed~\parencite{FS2006}: the continuum is augmented with an additional unknown field to be solved~\parencite{ARK+2022}. This is particularly true for second gradient theories, where a clear physical interpretation in terms of local deformed micro-trihedra is not immediately obvious, as it has been consistently addressed only recently~\parencite{FS2025}. Besides, even in approaches where generalised media are explicitly described by local moving frames, traditionally called \emph{directors of matter}~\parencite{For2006}, a conceptual ambiguity persists regarding their physical status :
\begin{itemize}
    \item Are they material parameters that explicitly encode the microstructural architecture (such as the primitive cell of a lattice) ?
    \item Or are they only kinematic descriptors (or \emph{gauges}) utilised by the experimenter to track the deformation of the meso-structure ?
\end{itemize}
In the geometric theory of dislocations~\parencite{Kro1968, YG2012}, these frames are tied to the crystal lattice where the explicit value of the frame matters. In pure elasticity, however, treating these frames as absolute structural descriptors undermines the objective of a general purposed continuum model, where the theory should remain independent of a specific microstructure. The difficulty of conceiving a general purpose elastic theory compatible with arbitrary micro-structures but maintaining an explicit micro-macro identification (in terms of kinematics, constitutive laws and energies) is an ongoing research topic~\parencite{LSS2026,RSI2026}. This distinction on the status of the directors of matter is thus essential for a consistent foundation of generalised continua and will be the main theoretical question addressed in this paper. Besides according to \parencite{IM2024}, other practical issues ranging from the formulation of consistent boundary conditions~\parencite{MGNM2016}, through the problem of the synthesis of micro-structure archieving a given property~\parencite{RSI2026}, to the experimental identification of constitutive laws~\parencite{FPF2024}, can be related to not always clear foundational definition of generalised media.

In hyperelasticity when conceiving constitutive law for generalised continua, the enforcement of objectivity (material frame-indifference) and the accounting of material symmetries are the essential conditions that a free energy density should verify~\parencite{TN1965}. Well understood in classical continuum mechanics, the choice of adequate objective strain measures~\parencite{Eri1999,La2022} and micro-inertia terms~\parencite{EE2007, MNA+2017, MN2018} for generalised continuum remains debated. The consistency of these invariance principles can be related to the requirement of general covariance within General Relativity~\parencite{Sou1958}. While classical objectivity emerges naturally as a limit of four-dimensional general covariance for standard continua~\parencite{KD2023}, recent findings show remarkable results for generalised media: second-grades theories may exhibit unexpected non-objective terms in their classical limit, \textit{i.e.} at infinite speed of light, but emerging from (relativistic) general covariant quantities~\parencite{CDDEK2025}.

To resolve these modeling ambiguities (classification and nature of models, status of the directors of matter, material frame indifference) and offer a unified framework to describe elastic mechanical models available in the literature, recent efforts have turned toward formulating rigorous geometric theories of generalised continua. However, existing geometric literature remains specialised on some topics and falls short of resolving some modeling difficulties for two reasons:
\begin{itemize}
    \item The majority of geometric frameworks are confined to modeling specific defect oriented phenomena like dislocations~\parencite{Laz2011, SY2021, NA2024, CCL2025} or the theory of inhomogeneities and material symmetries~\parencite{Nol2006, EE2007}. While very valuable, these studies focus heavily on material evolution and leave the elasticity of generalised media unaddressed;
    \item These geometric studies are typically conducted within the boundaries of a single pre-selected theory (\textit{e.g.}, Cosserat, micromorphic or strain-gradient). They do not seek to provide a unified classification capable of organising the available elastic models used in mechanics.
\end{itemize}
A promising framework to fulfill these objectives is gauge theory~\parencite{Ble1981}. While gauge theory is the mathematical cornerstone of the Standard Model in particle physics~\parencite{Ham2017}, its application in continuum mechanics has been restricted to some defect studies~\parencite{Laz2011}. From a mathematical point of view, gauge-theoretic methods provide a unified architecture for generalised continua. Gauge theories are formulated on principal bundles, of which the frame bundle (the collection of all pointwise frame on a manifold) is the archetypal example. Moving frames (local sections of the frame bundle) are widely used to model complex thin structures in finite strains~\parencite{SF1989, SS2020} which are considered as analogues of generalised continua for body of dimension lower than $3$. Besides, moving frames are also partly used in the geometric literature on generalised continua~\parencite{ED1998,JE2026}. Our goal is to use some of these concepts to provide a unified description of generalised sub-theories. Besides, a geometric approach can offer powerful tools that require particular mathematical structures: such as the Euler-Poincaré equations~\parencite{BR2017}, Hamiltonian formulations~\parencite{SMK1988,CL2026}, or robust variational integrators for computational mechanics~\parencite{CLS2024}.

% ---------------
\subsection*{Outline}
% --------------------

We begin in \cref{sec:Recal} by recalling the geometric formulation of classical Cauchy continua in statics and finite strains, emphasising the implementation of material frame-indifference via an invariance principle. Kinematics of generalised continua are introduced in \cref{sec:MovingFrames} by defining \emph{generalised configurations} as \emph{moving frames} over classical configurations. This rigorous definition, which slightly differ from other available definitions (\textit{e.g.} in \parencite{EE2007} or \parencite{YG2012}), allows for the derivation of the usual quantities of generalised continua without ad-hoc considerations. To match this perspective with usual mechanical models, \cref{sec:Defor} formalises the kinematic variables (\emph{deformations}) of the medium. Here, we contrast intrinsic tensor representations with coordinate-based formulations, establishing the definitions of \emph{generalised deformation}, \emph{micro-deformation}, and \emph{observed generalised deformation}. \cref{sec:FreeEn} introduces a first-gradient free energy functional and provides the mechanical justification for treating the observed generalised deformation as the fundamental variable of this energy.
The main theoretical contributions of this work are developed in the subsequent sections:
\begin{itemize}
    \item In \cref{sec:DiscExample}, we confront the non-physical dependency of the free energy on the reference generalised configuration by demonstrating that in generalised elasticity, the reference moving frame must be stripped of physical dependency to the micro-structure and treated as an arbitrary, user-defined kinematic descriptor, a \emph{choice of gauge} ;
    \item This physical argument of arbitrary generalised reference configuration is mathematically formalised in \cref{sec:FormalPrinc} ;
    \item Finally in \cref{sec:Gauge}, we establish that removing this dependency over the reference generalised configuration is a necessary and sufficient condition for enforcing this \emph{gauge-invariance} and recovers the usual micromorphic model. This result establishes that the micromorphic theory is a generalised purposed continuum theory describing the deformation of \emph{arbitrary chosen directors of matter}.
\end{itemize}
Following sections are dedicated to more practical, yet theoretical, results to address the problem of model classification raised in the introduction by providing an effective classification method. In \cref{sec:Classif}, the structural group reduction of the configuration space is introduced to obtain a systematic taxonomy of first-order generalised media. By reducing the structural group $\GL_3(\RR)$ to its closed subgroups, we identify the space of generalised configurations of a generalised continuum to a $\subG$-structure (based on the work of \parencite{MD2004} on material symmetries) providing a classification of sub-theories.

\cref{sec:strain} shows how strain-gradient continua naturally fit in our framework through convected frames~\parencite{BR2017,FS2025}. The lifting of the classical deformation, leading to the generalised deformation convecting frames, induces a natural \emph{compatible}/\emph{incompatible} decomposition of the micro-deformation in \cref{sec:decompMic} similar to the $\gradT^e\gradT^p$ decomposition of finite strain plasticity. This lifting also reveals to be the tool to define material frame indifference for generalised continua which is detailed in \cref{sec:MFI} and match the practice of the mechanical literature~\parencite{NF2007}. Finally in \cref{sec:constrain}, we propose a systematic Euclidean minimisation method to deduce second grade models (\emph{constrained media}) from first order media.

%\subsection*{Notations}
%As far as possible we adopt the following conventions:
%\begin{itemize}
%    \item Using uppercase letters $\Xref$ (or $0$ subscript) on \emph{reference configurations} and lowercase letters $\Xdef$ on \emph{deformed configurations};
%    \item Using bold italic font $\Xref$ for $3$D vector quantities and bold font $\gradT$ for matrices/order 2 tensors.
%\end{itemize}

% ------------------------------------------------------
\section{Classical continuum mechanics in finite strains}
\label{sec:Recal}
% ------------------------------------------------------

Classical non-linear continuum mechanics models a material by an abstract compact manifold with boundaries $\BB$ of dimension $d\in\{1,2,3\}$ called the \emph{body}~\parencite{TN1965,Ste2015,KD2021}. It represents the labels of the particles of the material and is equipped with a \emph{mass measure} $\mu$, a volume form on $\BB$. The body is embedded in the Euclidean space $\Euc$ by two \emph{placements} $\pref,\pdef$ (embeddings). The \emph{reference} $\coref$ and \emph{deformed configurations} $\codef$ are the images of the body by the placements and $\trans:=\pdef\circ\pref^{-1}$ is the \emph{deformation}.
\begin{equation*}
    \begin{mytikzcd}[ampersand replacement=\&]{diagCla}
        \& \BB \arrow[ld, bend right, "\pref", swap] \arrow[rd, bend left, "\pdef"] \& \\
        \coref\subset\Euc \arrow[rr, "\trans"] \& \& \codef\subset\Euc
    \end{mytikzcd}
\end{equation*}
A free energy $\EnTot$ with (specific) energy density $\EnLoc$ of first gradient in the deformation $\trans$ is usually stated to describe the deformation of the material from the reference to the deformed configuration. On the reference configuration (Lagrangian formulation) it gives,
\begin{equation}
    \label{eq:Energy$3$DCauchy}
    \EnTot_{\pref}[\trans]
    =
    \int_{\coref}
    \EnLoc\left(\Xref,\trans(\Xref),T_{\Xref} \trans=:\gradT_{\Xref}\right)
    \masref(\Xref) \vol_{\metRef}
\end{equation}
in which $\metRef$ is the Euclidean metric restricted to $\coref$ and $\masref$ the \emph{reference mass density} (defined by the push-forward of the mass measure $\mu$ by $\pref$).
%Thus, for any point of the reference configuration $\Xref\in\coref$, for any point $\Xdef\in\Euc$ (not necessarily coming from a classical deformation $\trans$) and any linear map $\theta_{\Xref,\Xdef}$ from $T_{\Xref}\coref$ to $T_{\Xdef}\Euc$ (not necessarily coming from the gradient of a deformation $\gradT$) the pointwise free energy of the material $\EnLoc(\Xref, \Xdef, \theta_{\Xref,\Xdef})$ must be well defined. Thus, the free energy density is a function on the \emph{space of $1$-jets} $J^1(\coref,\Euc)$ of maps from $\coref$ to $\Euc$ defined by the union of triplets~\parencite{Ble1981,Olv1995},
%\begin{equation}
%    \label{eq:defOneJetCla}
%    J^1(\coref,\Euc)
%    :=
%    \bigcup_{\substack{
%            \Xref\in\coref \\
%            \Xdef\in\Euc}
%    }
%    \left\{
%    (\Xref, \Xdef, \theta_{\Xref,\Xdef})
%    \,\Big\vert\,
%    \theta_{\Xref,\Xdef}:T_{\Xref}\coref\to T_{\Xdef}\Euc \text{ linear}
%    \right\}.
%\end{equation}
In the following, $j(\trans)$ denotes the map from a point $\Xref$ of the reference configuration $\coref$ to the triplet $(\Xref, \trans(\Xref), \gradT_{\Xref})$ for a deformation $\trans:\coref\to\codef$, \textit{i.e.}
\begin{equation}
    \label{eq:jetDefCla}
    j_{\Xref}(\trans)
    =
    (\Xref, \trans(\Xref), \gradT_{\Xref}=T_{\Xref}\trans),
    \quad \Xref\in\coref.
\end{equation}
Thus, the free energy density $\EnLoc$ must be well defined for any triplet $j_{\Xref}(\trans)$ coming from any deformation $\trans$ defined on any reference configuration $\coref$. The point-wise specific free energy of the system of \cref{eq:Energy$3$DCauchy} can be written compactly as $\EnLoc\left(j_{\Xref}(\trans)\right)$.
%The evaluation of the free energy density on a deformation $\trans$ at a point $\Xref$ is then given by $\EnLoc(j_{\Xref}(\trans))$.

%\begin{remark}
%    The space $J^1(\coref,\Euc)$ is a fiber bundle over the manifold $\coref$ and the jet $\Xref\mapsto j_{\Xref}(\trans)$ of a deformation $\trans:\coref\to\codef$ is a local section of this bundle over $\coref\subset\Euc$.
%\end{remark}

The notion of objectivity, material frame indifference and invariance by superposed rigid body motion, although often confused and controversial in the mechanical literature~\parencite{TN1965,Liu2005,Fre2009}, sought to formulate principles of covariance that has to be respected by constitutive laws. In its modern form~\parencite{YMO2006,Ste2007,Ste2015,KD2024}, it is formulated as the invariance of the free energy by the group of isometries of the Euclidean space,
\begin{equation}
    \label{eq:invCla}
    \EnTot_{\pref}[\diffI_*\trans]
    =
    \EnTot_{\pref}[\trans],
    \quad
    \forall \diffI\in\text{isometries of $\Euc$}.
\end{equation}
in which $_*$ denotes the action by push-forward given by,
\begin{equation}
    \label{eq:actionFuncCla}
    \diffI_*\trans := \diffI\circ\trans.
\end{equation}
Once a global coordinate system of the Euclidean space has been chosen (see \cref{ex:Sglob}), isometries express as
\begin{equation}
    \label{eq:defRigid}
    \diffI(\Xdef)
    :=
    \vtrans + \Rtrans\Xdef,
    \quad
    \vtrans\in\RR^3, \, \Rtrans\in\SO(3).
\end{equation}

%\todoBK{Pour avoir une expression en $\vtrans\in\RR^3$ et $\Rtrans\in\SO(3)$ il faut absolument un repère orthonormé. Je vois pas comment faire sinon (et si on fait un abus ça tue une partie de l'argumentaire que l'on donne plus tard).}

\begin{figure}[h]
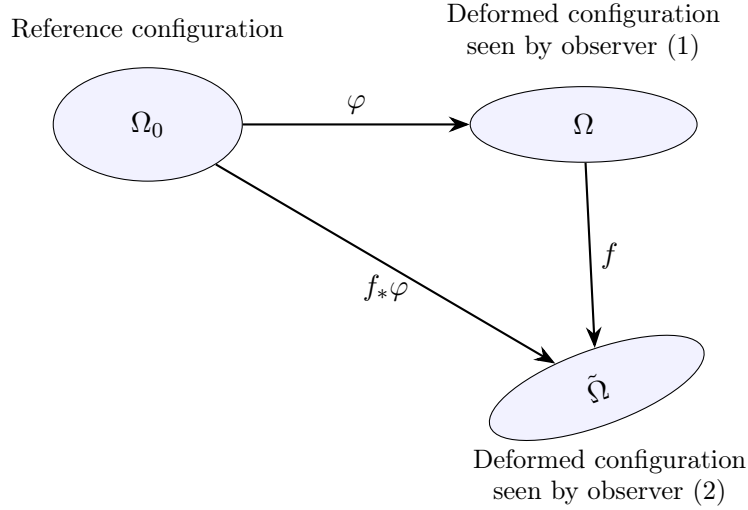

    \centering
    \tikzsetnextfilename{diagObj}
    \insertDiagObj
    \caption{Illustration of the invariance by isometries.}
    \label{fig:schObj}
\end{figure}

The invariance principle of \cref{eq:invCla} seeks to formulate the idea that two observers, related by an isometry, observing the ``same'' deformation $\trans$, should be able to compute the same free energy describing the system (see \cref{fig:schObj}).

Since $\diffI$ is a diffeomorphism of the Euclidean space, $\diffI_*\trans$ is a deformation with the same reference configuration $\coref$ and thus its free energy density $\EnLoc$ can be evaluated at a point $\Xref\in\coref$ by (using the notation introduced in \cref{eq:jetDefCla}),
\begin{equation}
    \label{eq:actionJetCla}
    \EnLoc\left(
    j_{\Xref}\left( \diffI_*\trans \right)
    \right)
    :=
    \EnLoc\left(
    \Xref, \diffI(\trans(\Xref)), T_{\trans(\Xref)}\diffI\circ\gradT_{\Xref}
    \right),
\end{equation}

%Since $\diffI$ is a diffeomorphism of the Euclidean space, it also acts on the triplet $j_{\Xref}(\trans):=(\Xref,\trans(\Xref),\gradT_{\Xref})$, given by the first \emph{jet} of a deformation $\trans$ (see \cref{eq:jetDefCla}), by the push-forward action $_*$,
%\begin{equation}
%    \label{eq:actionJetCla}
%    \diffI_* \left( \Xref, \trans(\Xref), \gradT_{\Xref} \right)
%    =
%    \left( \Xref, \diffI(\trans(\Xref)), T_{\trans(\Xref)}\diffI\circ\gradT_{\Xref} \right),
%\end{equation}
%in which the action on $\gradT_{\Xref}$ is the only one preserving the adequate tangent spaces, \textit{i.e.} mapping vectors of $T_{\Xref}\coref$ to vectors of $T_{\diffI(\trans(\Xref))}\Euc$.
%\todoBK{
%    Tu me dis d'être moins matheux mais je vois pas ce que ça apporte de détailler le prolongement des jets. Je peux à la limite mettre une phrase: "this is called the jet prolongation \parencite{Olv1995}".\\
%    Par ailleurs, tu m'as suggéré d'enlever les jets du papier mais certaines equations/arguments sont plus difficile à expliquer et je pense avoir garder leur utilisation au minimum.
%}
%It is straightforward to show the \emph{naturality} property relating the action of diffeomorphism on triplets $j_{\Xref}(\trans)$ (\cref{eq:actionJetCla}) and on a deformation $\trans:\coref\to\codef$ (see action defined in \cref{eq:actionFuncCla}) by,
%\begin{equation}
%    \label{eq:naturalityCla}
%    j_{\Xref}\left( \diffI_*\trans \right)
%    =
%    \diffI_*j_{\Xref}(\trans).
%    \quad
%    \forall\diffI\in\text{Diff}(\Euc),\,\Xref\in\coref.
%\end{equation}
The left hand side of the invariance principle of \cref{eq:invCla}, using the free energy defined in \cref{eq:Energy$3$DCauchy}, is given by, %(also using the naturality of \cref{eq:naturalityCla}),
\begin{equation*}
    \EnTot_{\pref}[\diffI_*\trans]
    =
    \int_{\coref} \EnLoc\left( j_{\Xref}(\diffI_*\trans) \right)
    \masref(\Xref) \vol_{\metRef}.
\end{equation*}
The right hand side of the invariance principle (\cref{eq:invCla}) is also an integration over $\coref$ with the same integration measure $(\pref)_*\mu=\masref\vol_{\metRef}$. Since the invariance principle must be true for any classical reference placement $\pref$, it must be true for arbitrary small classical reference configuration $\coref$. Therefore the invariance principle can be recast as a pointwise invariance for all $\Xref\in\coref$ on the free energy density for a deformation $\trans$ defined around this point by,
\begin{equation}
    \label{eq:invLocCla}
    \EnLoc\left( j_{\Xref}(\diffI_*\trans) \right)
    =
    \EnLoc\left( j_{\Xref}(\trans) \right),
    \quad
    \forall \diffI\in\text{isometries of $\Euc$}.
\end{equation}
%in which $\diffI_*$ is the push-forward by $\diffI$ (\cref{eq:actionFuncCla}).
%Since this equation must hold for any deformation $\trans$ and any reference configuration $\coref$, it imposes an invariance on the free energy density for all deformation $\trans$ defined  $(\Xref,\Xdef,\theta_{\Xref,\Xdef})$ around $\Xref\in\coref$ (\cref{eq:defOneJetCla}),
%\begin{equation*}
%    \EnLoc\left( \Xref, \diffI(\Xdef), T_{\Xdef}\diffI\circ\theta_{\Xref,\Xdef} \right)
%    =
%    \EnLoc\left( \Xref, \Xdef, \theta_{\Xref,\Xdef} \right),
%    \quad
%    \forall \diffI\in\text{sub-group of $\text{Diff}(\Euc)$}.
%\end{equation*}
For isometries , once a global coordinate system has been chosen on $\Euc$ (see \cref{eq:defRigid}), it reads (using \cref{eq:actionJetCla}),
\begin{equation}
    \label{eq:invJCla}
    \EnLoc\left( \Xref, \vtrans+\Rtrans\trans(\Xref), \Rtrans\circ\gradT_{\Xref} \right)
    =
    \EnLoc\left( \Xref, \trans(\Xref), \gradT_{\Xref} \right),
    \quad
    \forall\vtrans\in\RR^3,\,\forall\Rtrans\in\SO(3).
\end{equation}
It is the classical invariance imposed to energy densities and it is well-known that the free energy density $\EnLoc$ should only depends on the \emph{(right) Cauchy-Green tensor} $\CG=\gradT^\star\metDef\gradT$ and on the points $\Xref$ (accounting for inhomogeneities) ; $\gradT^\star$ is the non-metric transpose of the \emph{deformation gradient} $\gradT:=T\trans$ and $\metDef$ is the Euclidean metric restricted to $\codef$.

% ------------------------------------------------------
\section{Moving frames and generalised configurations}
\label{sec:MovingFrames}
% ------------------------------------------------------

The goal of this section is to rigorously define the configurations of a generalised continua (of order one) in a finite strain setting with clear mentions of the body in \cref{sec:genConfig}. This setting should be compatible with classical continuum mechanics and is based on the definitions of frames of the Euclidean space introduced in \cref{sec:movFrame}.

% -----------------
\subsection{Moving frames}
\label{sec:movFrame}
% -----------------

A \emph{pointwise frame} at a point $\Xdef$ of the $3$D Euclidean space $\Euc$ is primarily seen as a (not necessarily orthonormal) basis $(\LocFra)$ of the tangent space $T_{\Xdef}\Euc$. It is equivalently seen as a linear isomorphism $\fdef$ between $\RR^3$ and $T_{\Xdef}\Euc$. The equivalence is given by the image of the canonical basis $(\bm{e}_i)$ of $\RR^3$ (see \cref{fig:diagFrame}),
\begin{equation}
    \label{eq:frame}
    \LocFra
    =
    \fdef(\bm{e}_i).
\end{equation}
\begin{figure}[h]
    \centering
    \tikzsetnextfilename{diagFrame}
    \insertDiagFrame
    \caption{Diagram of a frame at $\Xdef\in\Euc$ as a basis $(\LocFra)$ and as an isomorphism $\fdef$.}
    \label{fig:diagFrame}
\end{figure}

The set of all pointwise frames at any point of the space $\FEuc$, called the \emph{frame bundle} of the Euclidean space, is defined by
\begin{equation}
    \label{eq:defFEuc}
    \FEuc
    :=
    \bigcup_{\Xdef\in\Euc} \{
    \fdef \text{ is a pointwise frame at } \Xdef\in\Euc
    \}.
\end{equation}

\begin{remark}
    \label{rem:2DEuc}
    In this work, the Euclidean space $\Euc$ is the three dimensional affine space but it can be replaced by the two dimensional Euclidean space to describe ``purely'' planar systems.
\end{remark}

\begin{remark}
    \label{rem:projection}
    The \emph{canonical projection} $\pi\colon\FEuc\to\Euc$ maps a frame $\fdef$ to its \emph{origin} $\Xdef$. Therefore $\pi^{-1}(\Xdef)$ is the set of all pointwise frames at $\Xdef$, called the \emph{fiber} above $\Xdef$. Given a domain $\codef$ of the Euclidean space, the set of all frames on this domain is denoted by $\FEuc_{\vert\codef}$, \textit{i.e.} the restriction of the frame bundle to the domain having $\codef$ as origins.
\end{remark}
A pointwise frame $\fdef$ at $\Xdef$, described by a basis $(\LocFra)$, can be transformed by an invertible matrix $\matA\in\GL_3(\RR)$ into another frame $\fdefp$ with the same origin $\Xdef$, described by a new basis $\LocFrap:=\fdef(\matA\bm{e}_i)$, \textit{i.e.}
\begin{equation}
    \label{eq:actGFrame}
    \LocFrap
    = \fdef(\matA\bm{e}_i)
    = (\fdef\circ\matA)(\bm{e_i})
    = \fdefp(\bm{e_i}).
\end{equation}
It is the (right) action of the \emph{structural group} $\GL_3(\RR)$ of real invertible $3\times 3$ matrices on the frame bundle $\FEuc$. This action is simply transitive at each fiber, meaning that for all points $\Xdef$ of the Euclidean space, two pointwise frame $\fdef,\fdefp$ at this point are related by a unique \emph{change of basis matrix} $\matA\in\GL_3(\RR)$ such that,
\begin{equation}
    \label{eq:frameMat}
    \fdefp
    =
    \fdef\circ\matA
    =:\fdef\matA.
\end{equation}

A \emph{moving frame} $\Sdef\colon\Xdef\mapsto\Sdef(\Xdef)$ is a smooth mapping that associates to each point $\Xdef$ of a domain of the Euclidean space a unique frame $\fdef$ at this point (see \cref{ex:Sglob}). For two moving frame $\Sdef,\Sdefp$ defined on a shared domain $\codef$, there exists a unique smooth \emph{transition function} $\matA:\codef\to\GL_3(\RR)$ (generalisation of \cref{eq:frameMat}) such that,
\begin{equation}
    \label{eq:transFunc}
    \Sdefp(\Xdef)
    =
    \Sdef(\Xdef)\matA(\Xdef),
    \quad \Xdef\in\codef,
\end{equation}
in which $\matA$ corresponds to the change of basis between the vectors describing $\Sdef(\Xdef)$ and those describing $\Sdefp(\Xdef)$.

\begin{example}[Canonical frame $\Sglob$ of $\Euc$]
    \label{ex:Sglob}
    On the affine Euclidean space $\Euc$ it is possible to choose a fixed and everywhere defined moving frame $\Sglob$. Indeed, choosing a point of the Euclidean space allows to identify $\Euc$ to a vector space in which a basis, typically orthonormal, can be chosen.
    %Given a frame of reference $R_{\Xdef_0}^{\mathrm{can}}$ at a point $\Xdef_0$ described by a basis $\left(\vv^{\mathrm{can}}_i\right)$, this frame can be translated to any point $\Xdef$ by keeping its basis fixed leading to a frame $\fdef^{\mathrm{can}}$. Therefore, choosing a frame of reference, typically orthonormal, allows to define a \emph{global} and ``fixed'' \emph{canonical moving frame} $\Sglob$ on $\Euc$.
\end{example}

With a choice of a global canonical frame $\Sglob$ on the Euclidean space (\cref{ex:Sglob}), the frame bundle $\FEuc$ can be \emph{trivialised}: any pointwise frame $\fdef$ can be uniquely identified by its origin $\Xdef$ (given by the canonical projection $\pi:\fdef\mapsto\Xdef$, see \cref{rem:projection}) and a $3\times 3$ real invertible matrix $\gdefp$, its \emph{matrix coordinates}, \textit{i.e.}
\begin{equation}
    \label{eq:trivialisation}
    \begin{array}{ccc}
        \FEuc & \to     & \Euc\times\GL_3(\RR) \\
        \fdef & \mapsto & \begin{pmatrix}
                              \Xdef \\ \gdefp
                          \end{pmatrix}
    \end{array}
\end{equation}
The matrix coordinates $\gdefp$ of a frame $\fdef$ is the change of basis matrix (or transition function) between the frame $\fdef$ and the canonical frame $\Sglob(\Xdef)$ at $\Xdef$, \textit{i.e.} $\fdef=\Sglob(\Xdef)\gdefp$ (\cref{eq:frameMat}).
\begin{remark}
    \label{rem:principalBundle}
    The frame bundle is the prototype of a $\subG$-principal fiber bundle, the geometric structure used in \emph{gauge theory}~\parencite{Ble1981,Ham2017}. In the vocabulary of gauge theory, a moving frame is a \emph{local section} or a \emph{local gauge choice} of the principal bundle and the action of invertible matrices on frames (\cref{eq:actGFrame}) is the \emph{action of the structural group} $\subG=\GL_3(\RR)$. The identification of a frame by its origin and a matrix coordinates is a \emph{trivialisation} of the frame bundle which is only local in general (but always possible), and global on the Euclidean space because it is an affine space (see \cref{ex:Sglob}). The frame bundle $\FMM$ of a manifold $\MM$ of dimension $d$ is defined in the same way as of the Euclidean space (\cref{eq:defFEuc}), but does not admit in general a global section (a canonical frame), \textit{i.e.}
    \begin{equation*}
        \FMM
        :=
        \bigcup_{m\in\MM} \left\{ \fmm:\RR^d\to T_m\MM \text{ isomorphism } \right\}.
    \end{equation*}
\end{remark}

% -----------------
\subsection{Generalised placements and configurations}
\label{sec:genConfig}
% -----------------

We define a \emph{generalised placement} $\Pdef$ as a smooth mapping of the body $\BB$ into the frame bundle $\FEuc$ such that its projection $\pdef:=\pi\circ\Pdef$ (see \cref{rem:projection}) is a placement (or embedding) of $\BB$ into the Euclidean space. It means that each particle of the body is associated to a pointwise frame such that the placement of the origins of the frames is a placement of $\BB$ into $\Euc$. We call $\Codef:=\Pdef(\BB)$ the \emph{generalised configuration}, $\Sdef:=\Pdef\circ\pdef^{-1}$ the \emph{associated moving frame} on $\codef:=\pdef(\BB)$ the \emph{classical configuration}.
\begin{equation}
    \label{eq:defAssociated}
    \begin{mytikzcd}[ampersand replacement=\&]{diagConfig}
        \&[3.5cm] \Codef\subset\FEuc \arrow[dd, "\pi", swap]
        \&[-1cm] \text{(Generalised configuration)}  \\
        \& \& \text{(Associated moving frame)} \\
        \text{(Body) }\BB \arrow[r, "\pdef\text{ (Classical placement)}", swap] \arrow[uur, "\Pdef\text{ (Generalised placement)}"]
        \& \codef\subset\Euc \arrow[uu, bend right, "\Sdef", swap]
        \& \text{(Classical configuration)}
    \end{mytikzcd}
\end{equation}
This generalised placement can be described using the trivialisation of \cref{eq:trivialisation} induced by a choice of canonical frame $\Sglob$ (see \cref{ex:Sglob}),
\begin{equation}
    \label{eq:placementTriv}
    \begin{array}{ccc}
        \BB & \to     & \Euc\times\GL_3(\RR)       \\
        \bb & \mapsto & \begin{pmatrix}
                            \Xdef := \pdef(\bb) \\
                            \gdefp := \Gdef(\pdef(\bb))
                        \end{pmatrix}
    \end{array}
\end{equation}
in which $\Xdef:=\pdef(\bb)$ is the origin of the frame $\Pdef(\bb)$ and $\Gdef:\codef\to\GL_3(\RR)$ is the matrix coordinates of the associated moving frame $\Sdef$ on $\codef:=\pdef(\BB)$, \textit{i.e.} the change of basis matrix between the frame $\Sdef(\Xdef)$ and the canonical frame $\Sglob(\Xdef)$. \cref{ex:rigidBody} and \cref{ex:Frenet} give possible generalised configurations for a body $\BB$ of dimensions $0$ (point) and $1$ (curve).
\begin{remark}
    \label{rem:smoothP}
    The condition of $\Pdef$ being smooth and that its projection is an embedding of $\BB$ into the Euclidean space is enough (for a compact body only) to insure that $\Pdef$ is an embedding of $\BB$ into the frame bundle $\FEuc$, and thus invertible on its image $\Codef$. However, the converse is not true, an embedding of $\BB$ into $\FEuc$ does not necessarily verifies that its projection is an embedding. Equivalently (for a compact body), a generalised placement $\Pdef$ can be constructed from a classical placement $\pdef$ of $\BB$ into $\Euc$ and a moving frame $\Sdef$ on $\codef:=\pdef(\BB)$ by $\Pdef:=\Sdef\circ\pdef$.
\end{remark}

\begin{figure}[h]
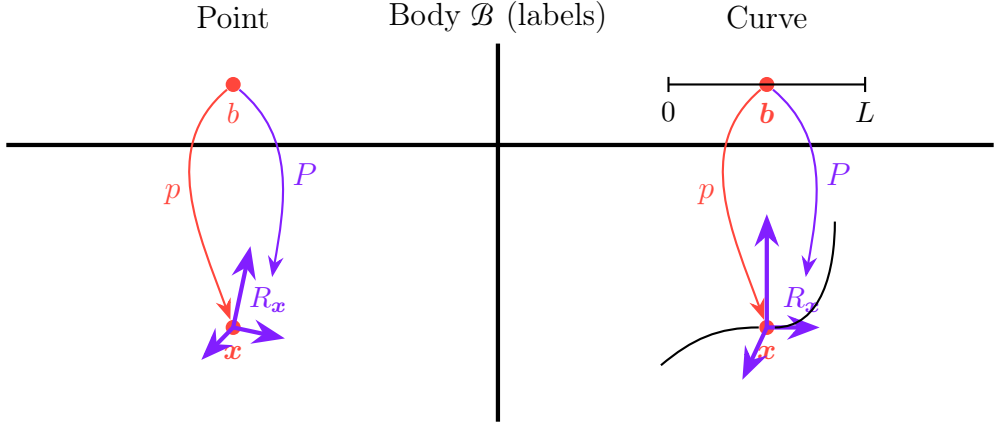

    \centering
    \insertDiagConfigEx
    \caption{Examples of generalised configurations for a point (\cref{ex:rigidBody}) and a curve (\cref{ex:Frenet}).}
    \label{fig:exampConfig}
\end{figure}

\begin{example}[Point, see \cref{fig:exampConfig}]
    \label{ex:rigidBody}
    A generalised configuration of a point, labelled by the singleton $\BB:=\{b\}$, is given by a frame $\Pdef(b)\in\FEuc$ of the Euclidean space. Using \cref{eq:placementTriv}, the origin of the frame $\Xdef:=\pdef(b)$ gives the position of the point and the directions of the frame is given by a the matrix $\gdefp:=\Gdef(\pdef(b))$ giving the coordinates of the frame in the canonical frame $\Sglob(\Xdef)$.
\end{example}

\begin{example}[Frame of a curve, see \cref{fig:exampConfig}]
    \label{ex:Frenet}
    Let $\codef\subset\Euc$ be a regular curve of the Euclidean space labelled by the segment $\BB:=[0,L]$, \textit{i.e.} placed in the Euclidean space by an embedding $\pdef$ ($\codef:=\pdef(\BB)$) with $\dot{\pdef}(b)\ne\bm{0}$ and $\ddot{\pdef}(b)\ne\bm{0}$. The non-orthonormalised trihedron,
    \begin{equation}
        \label{eq:Frenet}
        b\mapsto\left(
        \dot{\pdef}(b),\,\ddot{\pdef}(b),\, \dot{\pdef}(b)\times\ddot{\pdef}(b)
        \right)
    \end{equation}
    at a point $\Xdef:=\pdef(b)$, and the frame $\Pdef(b)$ associated to this trihedron (see \cref{eq:frame}), can be used as a generalised configuration $\Codef:=\Pdef(\BB)$ of the curve.
\end{example}

This definition is issued from~\parencite{CC1909} who define generalised configurations as a \emph{``$d$-parameters continuous distribution of trihedron''} with $d$ the dimension of the body considered. We reformulate this definition in a more contemporary language (intrinsic geometry rather than extrinsic), emphasising the importance of the body with the generalised placements and extending this definition to non-orthonormal trihedron. The main characteristics of the present definition are highlighted:
\begin{itemize}
    \item The state of a micromorphic media is described by a moving frame in which the origins of the frames represent its continuous macroscopic state and the frames represent the state of the microstructure at each macroscopic point which is in agreement with the usual mechanical theory~\parencite{Min1964,ES1964a};
    \item The body is used only as a labelling space, it does not carry additional mathematical structure nor physical characteristics unlike in recent geometrical theories~\parencite{EE2007,Laz2011,Ste2015,NA2024,CCL2025};
    \item Whatever the dimension of the body, the microscopic state is always described by a frame of the Euclidean space unlike in geometrical theories in which a frame of the configuration is used (for a two-dimensional body, the frame is composed by two tangent vector to the configuration). This choice is justified in accordance with the mechanical literature centered around thin structures in large strains in which moving frames are used~\parencite{BR2017,MS2025}.
\end{itemize}

\begin{remark}
    \label{rem:LinkEpstein}
    For a three dimensional body, our definition of generalised configurations and the definition used in \parencite{EE2007} (and \parencite{NA2024}) almost coincides. Indeed, our generalised placement $\Pdef:\BB\to\FEuc$ can be seen as a section over the pull-back bundle of the frame bundle over the body $\BB$. For a three dimensional body, this pull-back bundle and the frame bundle of the body (see \cref{rem:principalBundle}) are in bijection leading to almost the same modelling. The subtle difference is that the bijection requires to identify a moving frame on the body with a moving frame on a classical configuration which is itself a modelling hypothesis whose consequences will be made precise once the invariance principle is formulated, in \cref{sec:FormalPrinc} (more precisely in \cref{rem:repEpstein}).
\end{remark}

% ------------------------------------------------------
\section{Mapping frames and deformations}
\label{sec:Defor}
% ------------------------------------------------------

Similar to classical continuum mechanics, once configurations have been introduced, it is possible to define a deformation, mapping a reference configuration to a deformed one (see \cref{sec:Recal}). A generalised deformation, mapping generalised configurations made of frames, must satisfy certain properties to remain compatible with classical continuum mechanics. The \cref{sec:maps} presents those properties and consequences in a mathematical point of view and \cref{sec:genDef} details them from the mechanical perspective. In particular, it allows for the rigorous definition of the \emph{micro-deformation} $\tframe$, central in the formulation of generalised continua in the mechanical literature~\parencite{Ger1973}, as a natural constitutive quantity of the \emph{generalised deformation} $\Trans$.

% -----------------
\subsection{Maps between frames}
\label{sec:maps}
% -----------------

A \emph{fiber bundle isomorphism} $\Diff$ of the frame bundle is a smooth mapping between frames (belonging to $\mathrm{Diff}(\FEuc)$), such that for any frame $\fref\in\FEuc$,
\begin{equation}
    \label{eq:morphism}
    \Diff(\fref\matA)
    =
    \Diff(\fref)\matA,
    \quad
    \forall\matA\in \GL_3(\RR).
\end{equation}
Since any frame $\frefp$ at $\Xref:=\pi(\fref)$ can be written as $\fref\matA$ (\cref{eq:frameMat}), this \emph{equivariance} property means that $\Diff$ is completely determined by only one frame at each origin. In other words, the images by $\Diff$ of two frames with same origin $\Xref$ remains two frames with same origin that only differ from their initial relative change of basis. A fiber bundle isomorphism $\Diff$ sends an initial frame $\fref$ with origin $\Xref$ and matrix coordinates $\gref\in\GL_3(\RR)$ in a canonical frame $\Sglob$ (see \cref{eq:trivialisation}) to a frame $\Diff(\fref)$ with origin $\Xdef$ and matrix coordinates $\gdefp$, at first depending on both $\Xref$ and $\gref$, \textit{i.e.}
\begin{equation}
    \Diff\colon
    \begin{pmatrix}
        \Xref \\ \gref
    \end{pmatrix}
    \mapsto
    \begin{pmatrix}
        \Xdef(\Xref,\gref) \\ \gdefp(\Xref,\gref)
    \end{pmatrix}.
\end{equation}
Applying the projection $\pi$, retrieving the origins of the frame (see \cref{rem:projection}), to the equivariance property of \cref{eq:morphism} gives that the origin $\Xdef$ of the image only depends on $\Xref$ and we denote by $\trans\in\text{Diff}(\Euc)$ this mapping. To retrieve the matrix coordinates $\gdefp$, one has to apply the inverse of the canonical frame (see \cref{eq:trivialisation} which defines $\gref:=\Sglob(\Xref)^{-1}\circ\fref$) to the same equation to get,
\begin{equation*}
    \gdefp
    :=
    \left(
    \Sglob(\Xdef)^{-1} \circ \Diff(\Sglob(\Xref))
    \right)\gref.
\end{equation*}
Thus, by denoting $\Xref\mapsto\TframeX\in\GL_3(\RR)$ the map $\Xref\mapsto\Sglob(\trans(\Xref))^{-1} \circ \Diff(\Sglob(\Xref))$, a quantity that will be detailled in a few paragraphs, the fiber bundle isomorphism $\Diff$ can be expressed in a trivialisation induced by a canonical frame $\Sglob$ (\cref{eq:trivialisation}) by
\begin{equation}
    \label{eq:defTrivMorphism}
    \begin{array}{rccc}
        \Diff\colon & \Euc\times\GL_3(\RR)     & \to & \Euc\times\GL_3(\RR) \\
                    & \begin{pmatrix}
                          \Xref \\ \gref
                      \end{pmatrix}
                    & \mapsto
                    & \begin{pmatrix}
                          \Xdef := \diff(\Xref) \\
                          \gdefp := \TdiffX \gref
                      \end{pmatrix}
    \end{array}
\end{equation}
Given a moving frame $\Sref$ on a domain $\coref$ of the Euclidean space, the fiber bundle isomorphism $\Diff$ has a \emph{local form} $\Field:\coref\to\FEuc$ given by the pull-back of $\Trans$ by the moving frame $\Sref$, \textit{i.e.}
\begin{equation}
    \label{eq:defLocalForm}
    \Field(\Xref)
    :=
    \left( \Sref^*\Trans \right)(\Xref)
    =
    \Trans(\Sref(\Xref)).
\end{equation}
\begin{equation*}
    \begin{mytikzcd}[ampersand replacement=\&, column sep=5cm]{diagFieldDef}
        \FEuc \arrow[d, "\pi"] \arrow[r, "\Trans\text{ (Fiber bundle isomorphism)}"] \&  \FEuc \arrow[d, "\pi", swap] \\
        \coref \arrow[u, bend left, "\text{(Moving frame) }\Sref"] \arrow[ru, "\Field\text{ (Local form)}"] \arrow[r, "\trans\text{ (Origins mapping)}"] \& \Euc
    \end{mytikzcd}
\end{equation*}
Since $\Diff$ is a fiber bundle isomorphism, a local form $\Fieldp:\coref\to\FEuc$ of $\Diff$ in a different moving frame $\Srefp$ differing from $\Sref$ by a transition function $\matA:\coref\to\GL_3(\RR)$ (see \cref{eq:transFunc}) can be deduced from $\Field:=\Sref^*\Trans$ by a \emph{change of gauge formula},
\begin{equation}
    \label{eq:changeGaugeFormField}
    \Fieldp(\Xref)
    :=
    \Trans(\Srefp(\Xref))
    =
    \Trans(\Sref(\Xref)\matA(\Xref))
    =
    \Field(\Xref)\matA(\Xref).
\end{equation}
For a moving frame $\Sref$ on a domain $\coref$ described by its matrix coordinates $\Gref:\coref\to\GL_3(\RR)$ in a canonical frame $\Sglob$, the local form $\Field:=\Sref^*\Trans$ of the isomorphism $\Trans$ in the moving frame $\Sref$ is given in this trivialisation by (see \cref{eq:defTrivMorphism}),
\begin{equation}
    \label{eq:defTrivField}
    \Field\colon \Xref
    \mapsto
    \begin{pmatrix}
        \Xdef := \diff(\Xref) \\
        \gdefp := \TframeX\Gref(\Xref)
    \end{pmatrix}
\end{equation}

\begin{figure}[h]
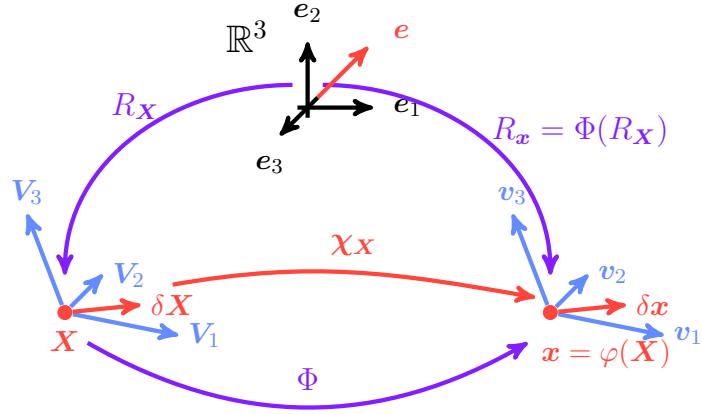

    \centering
    \tikzsetnextfilename{DiagTangent}
    \insertDiagTangent
    \caption{Illustration of the tangent space isomorphism $\tdiff_{\Xref}$ induced by a fiber bundle isomorphism $\Diff$.}
    \label{fig:DiagTangent}
\end{figure}

For each frame $\fref$, the frame $\Diff(\fref)$ induces a \emph{tangent space isomorphism} $\tdiff_{\Xref}$, since for all $\delta\Xref\in T_{\Xref}\Euc$ it exists a unique vector $\bm{e}$ such that $\delta\Xref=\fref(\bm{e})$,
\begin{equation}
    \label{eq:defTangentIsoVec}
    \delta\Xdef := \fdef(\bm{e})
    = \Diff(\fref)(\bm{e})
    = \Diff(\fref)(\fref^{-1}(\delta\Xref))
    =: \tdiff_{\Xref}\cdot\delta\Xref.
\end{equation}
This definition of $\tdiff_{\Xref}$ is independent of the choice of the frame $\fref$ because of the equivariance property of the isomorphism $\Diff$, making it an intrinsic quantity (see \cref{fig:DiagTangent} for illustration).

\begin{equation*}
    \begin{mytikzcd}[ampersand replacement=\&]{diagMicro}
        \& \RR^3 \arrow[dl, "\fref", bend right, swap] \arrow[dr, "\fdef:=\Diff(\fref)", bend left] \& \\
        T_{\Xref}\Euc \arrow[rr, "\tdiff_{\Xref}"] \& \& T_{\Xdef}\Euc
    \end{mytikzcd}
\end{equation*}

\begin{remark}
    \label{rem:microDefTrans}
    Looking at \cref{eq:defTangentIsoVec}, the \emph{tangent space isomorphism} $\tdiff_{\Xref}$ is determined by the isomorphism $\Diff$ and the frame $\fref$ at that point $\Xref$ and vice versa by,
    \begin{equation*}
        %\label{eq:defTangentIso}
        \tdiff_{\Xref}
        :=
        \Diff(\fref)\circ \fref^{-1}.
    \end{equation*}
\end{remark}

The tangent space isomorphism $\tdiff$ can be related to the mapping of the matrix coordinates $\TdiffX$ of $\Diff$ (see \cref{eq:defTrivMorphism}) by,
\begin{equation}
    \label{eq:matrixRepr}
    \TdiffX
    =
    \Sglob\left( \trans(\Xref) \right)^{-1}\circ \tdiff_{\Xref}\circ \Sglob(\Xref).
\end{equation}
Looking at this expression, the mapping of the matrix coordinates $\TdiffX$ is interpreted as the \emph{matrix representation} of the tangent space isomorphism $\tdiff_{\Xref}$ in the canonical frame $\Sglob$.

\begin{remark}
    \label{rem:fieldsEquiGauge}
    The data of the following three notions are equivalent : the fiber bundle isomorphism $\Diff$ ; the tangent space isomorphism $\tdiff$ ; the local forms $\Field$ in any moving frames with their change of gauge formula (\cref{eq:changeGaugeFormField}). This equivalence is a consequence of a central notion in gauge theory~\parencite{Ble1981,Ham2017}, the fiber bundle isomorphism $\Diff$ is an equivariant function with values in a manifold on which the structural group $\GL_3(\RR)$ acts (usually a vector space in gauge theory), the local forms $\Field$ on the base manifold $\Euc$ are characterising this function in local gauge choices $\Sref$ and the change of gauge formula characterises the equivariant nature of $\Diff$. Finally, the tangent space isomorphism $\tdiff$ should be interpreted as the \emph{matter field} used in physics. Note than in the case of the frame bundle, matter fields (sections of the associated bundle) interprets as tensor field on the base manifold which explains why $\tdiff$ is a two point tensor field. This equivalence will play a central role when interpreting the gauge invariance principle in \cref{sec:FormalPrinc}.
\end{remark}

However, the beginning of the next subsection will not only involve global mapping of the frame bundle (not even local mappings, \textit{i.e.} defined only on open sub-sets) but with smooth functions mapping two moving frames. Let $\trans:\coref\to\codef$ be a local diffeomorphism of the Euclidean space (similar to a classical configuration) and let $\Sref,\Sdef:\coref,\codef\to\FEuc$ be two moving frames on the domains $\coref$ and $\codef$. Let us consider the following mapping $\Transp$ between the two moving moving frames defined by,
\begin{equation}
    \label{eq:mappingMoving}
    \begin{array}{rcccl}
        \Transp\colon & \Sref(\coref) & \to     & \Sdef(\codef) &                                                        \\
                      & \fref         & \mapsto & \fdef         & := \Sdef\left( \trans\left( \pi(\fref) \right) \right)
    \end{array}
\end{equation}
Importantly, $\Transp$ can be constructed from the unique restriction to $\Sref(\coref)$ of a local (on $\FEuc_{\vert\coref}$, see \cref{rem:projection}) fiber bundle isomorphism $\Diff$ \parencite{ED1998}.

%\begin{remark}
%    \label{rem:constRest}
%    The set of fiber-preserving diffeomorphism that coincide with a mapping of moving frames of \cref{eq:mappingMoving} when restricted to the initial moving frame can be constructed from the set of automorphism of $\GL_3(\RR)$ such that the image of the identity $\Id$ is the identity~\parencite{ED1998}. For such an automorphism $h$, the following diffeomorphism restricted to $\Sref$ and $\Sdef$ coincides with the mapping $\Trans$,
%    \begin{equation*}
%        \fref          
%       \mapsto  
%       \Trans\left( \Sref(\Xref) \right)h\left( \fref\Sref(\Xref)^{-1} \right),
%        \quad
%        \text{with } \Xref:=\pi(\fref).
%    \end{equation*}
%\end{remark}

% --------------------
\subsection{Generalised deformations and micro-deformations}
\label{sec:genDef}
% ----------------------

Let $\Coref$ and $\Codef$ be two generalised configurations called the \emph{generalised reference} and \emph{deformed configurations}. All the previously introduced spaces and mappings related to generalised configurations will be denoted by a subscript $0$ when referring to the generalised reference configuration $\Coref$: the generalised placement $\Pref$; the classical placement $\pref:=\pi\circ\Pref$, the associated moving frame $\Sref:=\Pref\circ\pref^{-1}$ and the classical configuration $\coref:=\pref(\BB)$. Elements of those sets will be denoted by capital letters or $0$ subscripts: $\Xref$ for points of the classical configuration and $\gref$ for matrix coordinates of frames of the generalised configuration.\\

Similar to classical continuum mechanics, since the generalised placements $\Pref,\Pdef$ are bijective on their images $\Coref,\Codef$ (see \cref{rem:smoothP}), we define the \emph{generalised deformation} $\Trans$ as $\Trans:=\Pdef\circ\Pref^{-1}$. The generalised deformation $\Trans$ contains the data of the classical deformation $\trans\colon\coref\to\codef$ because it contains the deformation of the origin of the frames. Indeed, since $\pref$ and $\pdef$ are the placements of the body into the Euclidean space associated to the generalised placements $\Pref$ and $\Pdef$, we can define the classical deformation $\trans:=\pdef\circ\pref^{-1}$ mapping $\coref$ to $\codef$ (see \cref{ex:cubeConfig}). We define the \emph{observed generalised deformation} $\Field:=\Trans\circ\Sref$.
\begin{equation*}
    \begin{mytikzcd}[ampersand replacement=\&, column sep=6cm]{diagDef}
        \Coref \arrow[r, "\Trans\text{ (Generalised deformation)}"] \arrow[d, "\pi"]
        \& \Codef \arrow[d, "\pi", swap] \\
        \coref \arrow[u, bend left, "\Sref"] \arrow[r, "\trans\text{ (Classical deformation)}"] \arrow[ur, "\Field\text{ (Observed gen. def.)}"]
        \& \codef \arrow[u, bend right, "\Sdef", swap]
    \end{mytikzcd}
\end{equation*}

In terms of mechanical modelling, since the state of a micromorphic media is described by a moving frame, the generalised deformation of a micromorphic media is how those moving frames deform both in terms of origins (the macroscopic state) and in terms of directions (the microscopic state) as described in the mechanical and geometrical theory~\parencite{Eri1999,ED1998}.

\begin{figure}[h]
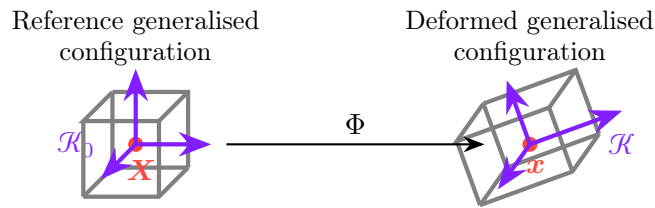

    \centering
    \tikzsetnextfilename{diagCube}
    \insertDiagCube
    \caption{Diagram showing a uniform deformation of a cube identified as a $0$ dimensional generalised deformation $\Trans$ (see \cref{ex:cubeConfig}).}
    \label{fig:DiagCube}
\end{figure}

\begin{example}[Small cube, see \cref{fig:DiagCube}]
    \label{ex:cubeConfig}
    Let us consider a small cube in the Euclidean space, in its reference state the center of the cube $\Xref$ is the classical reference configuration $\coref$ and in order to follow a deformation of the cube a trihedron is drawn on it (a choice of frame $\fref$) representing the reference generalised configuration $\Coref$ (similar to \cref{ex:rigidBody}). Therefore a uniform deformation (with a possible rigid body motion) of the cube will move its center as well as the trihedron. The new position of the center of the cube is $\Xdef$, the classical deformed configuration $\codef$, and the new directions of the trihedron is a frame $\fdef$, the generalised deformed configuration $\Codef$. The mapping from $\Coref$ to $\Codef$ gives a $0$ dimensional generalised deformation $\Trans$.
\end{example}

\begin{figure}[h]
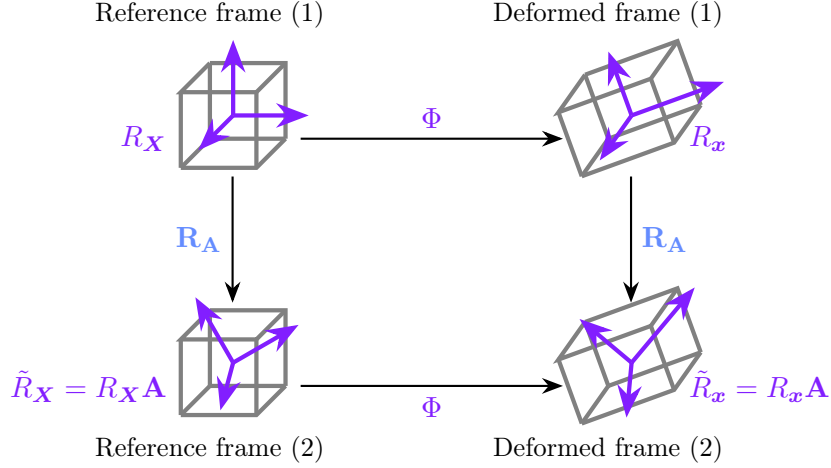

    \centering
    \insertDiagExtension
    \caption{Diagram illustrating the extension of a generalised deformation in which $\act_{\matA}$ denotes the right multiplication by a matrix $\matA\in\GL_3(\RR)$.}
    \label{fig:DiagMorphism}
\end{figure}

As described in the previous subsection (see \cref{eq:mappingMoving}), the generalised deformation $\Trans:=\Pdef\circ\Pref^{-1}$, initially defined on the generalised configuration $\Coref$, can be uniquely extended to a fiber bundle isomorphism (\cref{eq:morphism}) which by abuse of notations will also be denoted by $\Trans$. At this stage, the consequence of this extension can be illustrated pointwisely: given a frame $\fref$ and its image $\fdef:=\Trans(\fref)$ by the generalised deformation, applying the ``same'' generalised deformation $\Trans$ to a twisted reference frame $\frefp:=\fref\matA$, sharing the same origin, yields a twisted deformed frame $\fdefp:=\fdef\matA$, differing from $\fdef$ by the same initial transformation $\matA\in\GL_3(\RR)$ (see \cref{fig:DiagMorphism}).

\begin{remark}
    \label{rem:extensionMorphi}
    Extending the generalised configuration to a fiber bundle isomorphism is not only a mathematical convenience, it will become clear once the invariance principle of \cref{sec:Princ} is explored in \cref{sec:Gauge}. It seems that this extension is the only choice compatible, both mathematically and physically, with that principle.
\end{remark}

\begin{figure}[h]
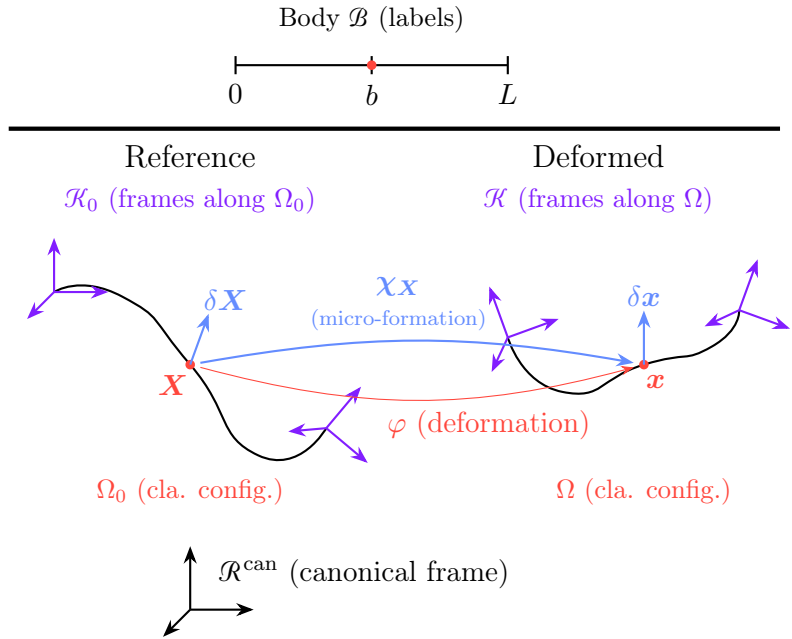

    \centering
    \insertDiagConfigMic
    \caption{Diagram of a one-dimensional micromorphic media with the micro-deformation.}
    \label{fig:Diag1DMic}
\end{figure}

The generalised deformation $\Trans$, which is now considered to be a fiber bundle isomorphism (\cref{eq:morphism}), induces a mapping of tangent spaces $\tframe$ between the classical reference and deformed configurations (see \cref{eq:defTangentIsoVec}): the \emph{micro-deformation} used in the mechanical literature. Taking $\bm{e}$ as one of the basis vector $(\bm{e}_i)$ of $\RR^3$ in \cref{rem:microDefTrans}, the micro-deformation then maps the vectors associated to the frame $\fref$ to the one associated to $\Trans(\fref)$ (see \cref{fig:Diag1DMic}) as described in \parencite{Eri1999}. The micro-deformation is defined geometricaly as a tangent space isomorphism by (see \cref{eq:defTangentIsoVec}),
\begin{equation*}
    \tframe_{\Xref}\cdot\delta\Xref
    :=
    \Trans(\fref)\left( \fref^{-1}(\delta\Xref) \right),
    \quad
    \delta\Xref\in T_{\Xref}\Euc.
\end{equation*}

\begin{remark}
    For a three dimensional media ($\text{dim}\,\BB=3$), the micro-deformation $\tframe_{\Xref}$ and the gradient of the classical deformation $\gradT_{\Xref}=T_{\Xref}\trans$ are of same nature: tangent space isomorphisms from $T_{\Xref}\coref\cong T_{\Xref}\Euc$ to $T_{\Xdef}\codef\cong T_{\Xdef}\Euc$, \textit{i.e.} $\gradT$ and $\tframe$ are two points tensors with same source and target points. For a body of lower dimension the micro-deformation remains a isomorphism between three dimensional spaces ($T_{\Xref}\Euc$ to $T_{\Xdef}\Euc$) whereas the gradient of the deformation is only defined from $T_{\Xref}\coref$ to $T_{\Xdef}\codef$ which are of same dimension of the body.
\end{remark}

As described previously, using the trivialisation of fiber bundle isomorphisms of \cref{eq:defTrivMorphism}, we denote by $\Tframe$ the \emph{matrix representation} of the micro-deformation $\tframe$ in the canonical frame $\Sglob$. It is obtained for any frame $\fref\cong(\Xref,\gref)$ on the classical reference configuration which image by the generalised deformation is $\Trans(\Xref)\cong(\Xdef,\gdefp)$, thus
\begin{equation}
    \label{eq:MicroMatTriv}
    \Trans\colon \begin{pmatrix}
        \Xref \\ \gref
    \end{pmatrix}
    \mapsto
    \begin{pmatrix}
        \Xdef  = \trans(\Xref) \\
        \gdefp  := \TframeX \gref
    \end{pmatrix}
\end{equation}
Similarly (see \cref{eq:defTrivField}), the observed generalised deformation $\Field$ in the reference associated moving frame $\Sref$ described by its matrix coordinates $\Gref:\coref\to\GL_3(\RR)$ can be expressed as,
\begin{equation}
    \label{eq:FieldTriv}
    \Field\colon \Xref
    \mapsto
    \begin{pmatrix}
        \Xdef  = \trans(\Xref) \\
        \gdefp  := \TframeX \Gref(\Xref)
    \end{pmatrix}
\end{equation}
Notice that the matrix representation $\TframeX$ of the micro-deformation depends implicitly on the classical deformation but solely on $\Xref$ and not on $\gref$ since the micro-deformation $\tframe_{\Xref}$ only depends on $\Xref$ (see \cref{eq:defTangentIsoVec}). Indeed, its matrix representation is given by (see \cref{eq:matrixRepr}),
\begin{equation}
    \label{eq:MatrixRepr}
    \TframeX
    =
    \Sglob\left(\trans(\Xref)\right)^{-1} \circ \tframe_{\Xref} \circ \Sglob(\Xref).
\end{equation}
In the mechanical literature, the micro-deformation $\tframe$ and its matrix representation $\Tframe$ are not distinguished and the generalised deformation $\Trans$, nor its local form $\Field$, are introduced. While the data of the micro-deformation $\tframe$ is equivalent to the data of the generalised deformation $\Trans$, the data of its matrix representation $\Tframe$ is not sufficient to reconstruct the generalised deformation, one needs the classical deformation $\trans$. It is a similar separation with the gradient of the deformation $\gradT$ which, mathematically contains the information of the mapping $\trans$, but its matrix representation in a frame does not (see \cref{rem:Fmatrix}).

\begin{remark}
    \label{rem:Fmatrix}
    The matrix representation of the gradient of the deformation $\gradT$ is given in $3$D by (identical to the micro-deformation \cref{eq:MatrixRepr}),
    \begin{equation}
        \label{eq:defGradMat}
        \GL_3(\RR)\ni[\gradT_{\Xref}]
        :=
        \Sglob(\trans(\Xref))^{-1}\circ\gradT_{\Xref}\circ\Sglob(\Xref).
    \end{equation}
    For a body of dimension lower than $3$, frames with vectors belonging to the tangent space of the configurations (such as $(\partial_{x^i})$ with $x^i$ a coordinate system on the configuration) must be used which a priori are not the case for the canonical frame $\Sglob$.
\end{remark}

In this section we have established the key kinematic quantities necessary for describing a change of generalised configurations. We recover the classical mechanical quantities with the deformation $\trans$ and its gradient $\gradT$. Besides, we recover the essential quantity related to generalised continua with the micro-deformation $\tframe$. The additional data carried by our framework is the distinction between the micro-deformation and its matrix representation $\Tframe$ and its correspondence with the generalised deformation $\Trans$ used in geometric theories~\parencite{EE2007}. We also introduced a new quantity, the observed generalised deformation $\Field$, that will be key in the next section and emphasised the mathematical equivalence between the three quantities $\Trans,\tframe$ and $\Field$ in \cref{rem:fieldsEquiGauge}. Finally, our contribution regarding the geometric foundation of generalised continua is that those definitions are also valid for body with dimension lower than $3$ thanks to the adopted definitions of generalised configurations.

% ------------------------------------------------------
\section{Free energy associated to a generalised deformation}%
\label{sec:FreeEn}
% ------------------------------------------------------

Similar to classical continuum mechanics, one wants to state a free energy $\EnTot$ with specific free energy density $\EnLoc$ to describe the ``deformation'' of the continuous media from a generalised configuration $\Coref$, to the generalised configuration $\Codef$. Mechanically, the free energy is evaluated at each macroscopic point $\Xref$ of the classical reference configuration $\coref$ and is associated to the deformation of the reference frame $\Sref(\Xref)$ at this point to the deformed frame $\Trans(\Sref(\Xref))$. Therefore, the free energy density in its most general form depends on the reference moving frame $\Sref$ and on the first gradient (for a first gradient theory) of the observed generalised deformation $\Field=\Trans\circ\Sref$ (more details in \cref{sec:Gauge}).

\begin{equation*}
    \begin{mytikzcd}[ampersand replacement=\&]{diagField}
        \Coref \arrow[d, "\pi"] \arrow[r, "\Trans"] \& \Codef \arrow[d, "\pi", swap] \\
        \coref \arrow[u, bend left, "\Sref"] \arrow[ru, "\Field"] \arrow[r, "\trans"] \& \codef \arrow[u, bend right, "\Sdef", swap]
    \end{mytikzcd}
\end{equation*}

The formulation in terms of the observed generalised deformation $\Field$ with an integration over the classical configuration $\coref$ is given by,
\begin{equation}
    \label{eq:EnTotField}
    \EnTot_{\pref,\Sref}[\Field]
    =
    \int_{\coref} \EnLoc\left(
    \Sref(\Xref), \Field(\Xref), T_{\Xref}\Field
    \right) \masref(\Xref) \vol_{\metRef},
\end{equation}
in which $\Sref:\coref\to\Coref$ is the associated moving frame to the generalised placement $\Pref:=\Sref\circ\pref$ (\cref{eq:defAssociated}) and the mass density $\masref$ is implicitly given by the placement $\pref$ (see \cref{sec:Recal}).\\

With the trivialisation of the frame bundle (\cref{eq:trivialisation}), the associated moving frame $\Xref\mapsto\Sref(\Xref)$ (to the generalised configuration $\Coref$) is entirely determined by its origin $\Xref$ and its matrix coordinates $\Gref:\coref\to\GL_3(\RR)$. Similarly (\cref{eq:FieldTriv}), the observed generalised deformation $\Field$ is entirely determined by the classical deformation $\trans:\coref\to\codef$ mapping the origins of the frames and the matrix representation of the micro-deformation $\Xref\mapsto\TframeX\in\GL_3(\RR)$ mapping the matrix coordinates of the frames. Therefore, the observed generalised deformation $\Field$ is described in a trivialisation (using a canonical frame $\Sglob$) by \cref{eq:FieldTriv}, \textit{i.e.}
\begin{equation}
    \label{eq:recallTrivField}
    \Field\colon \Xref
    \mapsto
    \begin{pmatrix}
        \Xdef  = \trans(\Xref) \\
        \gdefp  := \TframeX \Gref(\Xref)
    \end{pmatrix}
\end{equation}

Let $s\mapsto\Xref(s)$ be a smooth path of points in the classical reference configuration $\coref$ with $\Xref:=\Xref(0)$ and $\delta\Xref:=\dot{\Xref}(0)$. The image of this path of points by the observed generalised deformation $\Field$ induces a path of frames in $\Codef$ described by a path of origins $s\mapsto\Xdef(s)$ and a path of matrix coordinates $s\mapsto\gdefp(s)$ given by the trivialisation of $\Field$ recalled in \cref{eq:recallTrivField},
\begin{equation}
    s\mapsto
    \begin{pmatrix}
        \Xdef(s)  = \trans(\Xref(s)) \\
        \gdefp(s)  := [\tframe_{\Xref(s)}] \Gref(\Xref(s))
    \end{pmatrix}
\end{equation}
Consequently, the derivative of the observed generalised transformation $\Field$ at $\Xref$ in the direction $\delta\Xref\in T_{\Xref}\coref$ is given in a trivialisation by,
\begin{equation}
    \label{eq:derivativeField}
    T_{\Xref}\Field\cdot\delta\Xref
    =
    \frac{\dd}{\dd s} \begin{pmatrix}
        \Xdef(s) \\
        \gdefp(s)
    \end{pmatrix}
    =
    \begin{pmatrix}
        \gradT_{\Xref}\cdot\delta\Xref \\
        \left( T_{\Xref}\Tframe\cdot\delta\Xref \right) \Gref(\Xref) + \TframeX\left( T_{\Xref}\Gref\cdot\delta\Xref \right)
    \end{pmatrix}
    =:
    \begin{pmatrix}
        \delta\Xdef \\
        \delta\gdefp
    \end{pmatrix},
\end{equation}
in which $\delta\Xdef\in T_{\Xdef}\coref$ is the resulting origin (\emph{horizontal}) variation and $\delta\gdefp\in T_{\gdefp}\GL_3(\RR)$ the resulting matrix coordinates (\emph{vertical}) variation. The differential $T_{\Xref}\Field\cdot\delta\Xref$ of the observed generalised deformation in a direction $\delta\Xref$ is characterised by its horizontal variation $\delta\Xdef$ given by the gradient of the deformation $\gradT_{\Xref}$ and a vertical variation $\delta\gdefp$ given by,
\begin{equation}
    \label{eq:initialDeriv}
    D^V_{(\Xref,\Gref)}\Tframe\cdot\delta\Xref
    :=
    \left(
    \act_{\Gref(\Xref)}\circ T_{\Xref}\Tframe + \TframeX T_{\Xref}\Gref
    \right)\cdot\delta\Xref,
\end{equation}
in which $\act_{\matA}$ denotes the right multiplication by the matrix $\matA\in\GL_3(\RR)$. Finally, the free energy of the system from \cref{eq:EnTotField} can be expressed in a trivialisation given by a canonical frame $\Sglob$ as only depending on the classical deformation $\trans$ and the micro-deformation $\tframe$ (using its matrix representation $\Tframe$ in $\Sglob$),
\begin{equation*}
    %\label{eq:EnTotCTri}
    \EnTotC_{\pref,\Gref}[\trans,\tframe]
    =
    \int_{\coref} \EnLocC\left(
    \Xref, \Gref(\Xref),
    \trans(\Xref), \TframeX \Gref(\Xref),
    \gradT_{\Xref},
    D^V_{(\Xref,\Gref)}\Tframe
    \right) \masref(\Xref) \vol_{\metRef},
\end{equation*}
in which $\EnTotC$ with density $\EnLocC$ corresponds to the free energy and the free energy density described in this trivialisation. To simplify computations and arguments in the next section, it will be easier to consider a more general free energy density $\EnLocC$ depending separately on $\TframeX$ (since $\Gref(\Xref)$ is invertible and among the arguments of $\EnLocC$), $T_{\Xref}\Tframe$ and $T_{\Xref}\Gref$ (using similar arguments), \textit{i.e.}
\begin{equation}
    \label{eq:EnTotCTri}
    \EnTotC_{\pref,\Gref}[\trans,\tframe]
    =
    \int_{\coref} \EnLocC\left(
    \Xref, \Gref(\Xref),
    \trans(\Xref), \TframeX,
    \gradT_{\Xref}, T_{\Xref}\Tframe,
    T_{\Xref}\Gref
    \right) \masref(\Xref) \vol_{\metRef}.
\end{equation}
In the following, similarly to classical continuum mechanics (see \cref{eq:jetDefCla}), for a deformation $\trans$, a micro-deformation $\tframe$ and matrix coordinates of a reference moving frame $\Gref$ defined on a classical reference configuration $\coref$, $j(\trans,\tframe,\Gref)$ will denote the map from a point $\Xref\in\coref$ to the $7$-uplet,
\begin{equation}
    \label{eq:jetDefC}
    j_{\Xref}(\trans,\tframe,\Gref)
    :=
    (\Xref, \Gref(\Xref),
    \trans(\Xref), \TframeX,
    \gradT_{\Xref}, T_{\Xref}\Tframe,
    T_{\Xref}\Gref).
\end{equation}

%\begin{remark}
%    If one wanted to develop a theory in which the reference moving frame (through $\Gref$) would act as a physical description of matter (in the same manner as $\Xref$), the free energy density would be with values in $\Euc\times\GL_3(\RR)$. As discussed in \cref{sec:Princ} this is not our point of view and every values of $\gref$ is possible because we supposed that the generalised deformation is a fiber bundle isomorphism, \textit{i.e.} defined for all frames on the reference configuration.
%\end{remark}

%The evaluation of the trivialised free energy density $\EnLocC$ of \cref{eq:EnTotCTri} on a deformation $\trans$ and a micro-deformation $\tframe$ for a generalised reference configuration described by its matrix coordinates $\Gref$ at a point $\Xref$ is then given by $\EnLocC(j_{\Xref}(\trans,\tframe,\Gref))$. 
Thus, the free energy density $\EnLocC$ must be well defined for any $7$-uplet $j_{\Xref}(\trans,\tframe,\Gref)$ coming from any deformation $\trans$, any micro-deformation $\tframe$ and matrix coordinates of a reference frame $\Gref$, defined on any reference configuration $\coref$. The trivialised free energy of the system defined in \cref{eq:EnTotCTri} expresses compactly as,
\begin{equation}
    \label{eq:EnTotCTriCompa}
    \EnTotC_{\pref,\Gref}[\trans,\tframe]
    =
    \int_{\coref} \EnLocC(j_{\Xref}(\trans,\tframe,\Gref))
    \masref(\Xref) \vol_{\metRef}.
\end{equation}

\begin{remark}
    The above equation (\cref{eq:EnTotCTriCompa}) also expresses by forgetting some dependencies on $\Xref$ as,
    \begin{equation*}
        \EnTotC_{\pref,\Gref}[\trans,\tframe]
        =
        \int_{\coref} \EnLocC\left(
        \Xref, \Gref,
        \trans, \Tframe,
        \gradT, T\Tframe,
        T\Gref
        \right) \masref \vol_{\metRef}.
    \end{equation*}
\end{remark}

% ------------------------------------------------------
\section{Towards a missing invariance principle ?}%
\label{sec:Princ}
% ------------------------------------------------------

The explicit dependency of the trivialised free energy density $\EnLocC$ on the matrix coordinates of the reference moving frame $\Gref$ and its gradient $T\Gref$ in \cref{eq:EnTotCTri} represents a fundamental difference with usual formulations of micromorphic media~\parencite{For2006}. Mathematically, this dependency implies that the reference frame at each point $\fref$ (given by $\Gref(\Xref)$) encodes an absolute state of the microstructure. Physically, it means that the pointwise energetic cost of deforming an initial frame $\fref := \Sref(\Xref)$ to a deformed frame $\fdef := \tframe_{\Xref} \circ \fref$ (see \cref{rem:microDefTrans}) is not solely a function of the micro-deformation $\tframe$, but is conditioned by the precise values of $\fref$ and $\fdef$. Without further restrictions, such a formulation implies that it is impossible to relate the energy of identical relative changes (same micro-deformation) of ``micro-states'' if they originate from a different reference frame. The reference moving frame would act as a variable in the same sense that the point $\Xref$ can encode inhomogeneities in the material properties.

% -------------------------------------
\subsection{Discussion over examples}
\label{sec:DiscExample}
% -------------------------------------

To recover the usual formulation of micromorphic media, we state that the choice of the reference moving frame is an arbitrary convention required for kinematic measurements, rather than an intrinsic descriptor of material architecture/meso-structure. To clarify the potential status of a reference frame $\fref$, it is instructive to see its possible interpretation in two examples (\cref{ex:gaugeFrame} and \cref{ex:materialFrame}) considering similar periodic $2$D truss like micro-structured materials that are frequently homogenised into generalised continua. Previous works have shown that under elastic loading, these lattices can be homogenised into a continuous generalised theory using micro-macro energetic identification~\parencite{AMI2017}. In these examples, the different scales (see \cref{fig:schScale}) are identified as follow:
\begin{itemize}
    \item The macroscopic scale corresponds to the position of the node of the lattice, which when the cells are considered sufficiently small, can be identified as a continuous domain ;
    \item The mesoscopic scale corresponds to the arrangement of the lattice composed of the beams, the rotation of the beams (which will be identified to the Cosserat rotation) are the kinematic descriptors of this scale ;
    \item The microscopic scale correspond to the behaviour of the material composing the beams modelled by homogenous Euler-Bernoulli beams.
\end{itemize}

\begin{figure}[h]
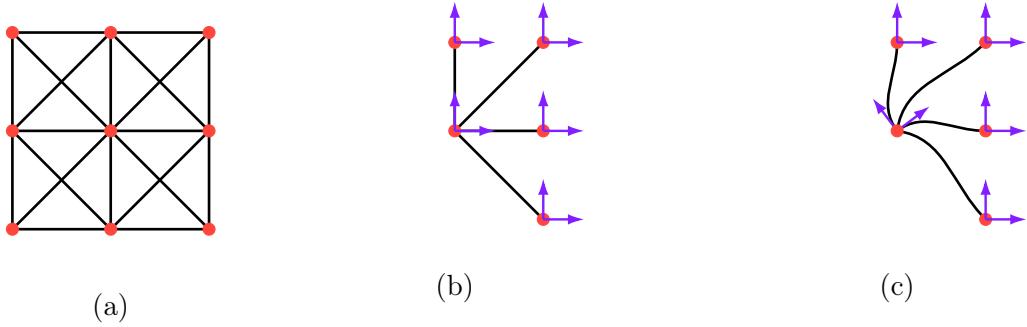

    \centering
    \insertDiagBeams
    \caption{(a) Square lattice connected by Euler-Bernoulli beams (similar to \parencite{AC1968}). (b) Initial configuration of a primitive cell with a choice of directors (purple) at each node tangent to the vertical and horizontal beams. (c) Deformed configuration with no node displacements but with the rotation of the center node. \parencite{AC1968} showed that the rotation of the node (described by the rotation of the frame attached to the node) is related, when homogenisation is performed, to the Cosserat degree of freedom.}
    \label{fig:BeamLattice}
\end{figure}

\begin{example}[see \cref{fig:BeamLattice}]
    \label{ex:gaugeFrame}
    Consider the classical homogenisation of a $2$D periodic square lattice of Euler-Bernoulli beams into a Cosserat continuum~\parencite{AC1968,CLH+2014}. Under elastic loadings, the micro-rotation (the Cosserat degree of freedom) is energetically identified with the localised curvature of the beams at each lattice node. To track this rotation, a set of directors is introduced. Structurally, it is convenient to align these directors with the tangents of two orthogonal beams at one node. However, this alignment is purely a matter of mathematical convenience. Any other initial directors, even some varying arbitrarily from node to node, remains equally valid since their relative evolutions would track the deformation of the truss. Just as a reference coordinate system is arbitrarily chosen to compute macro-displacements, the reference directors act as a spatial \emph{gauge} to measure micro-rotations. The elastic energy stored in the bending beams depends on the relative change of state, and must remains invariant under an arbitrary change of this reference gauge.
\end{example}

\begin{figure}[h]
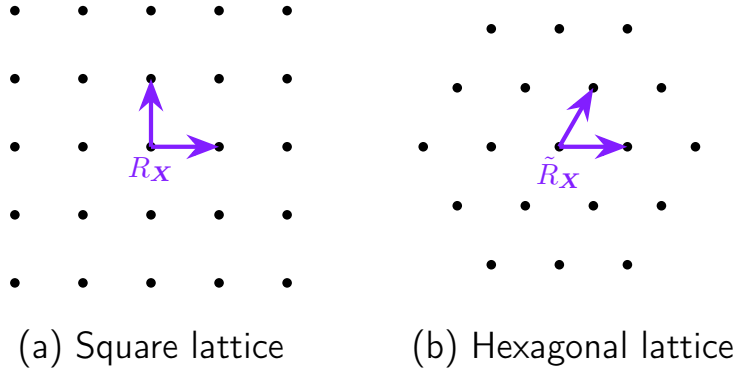

    \centering
    \insertDiagLattice
    \caption{Lattices identified by primitive translation frames.}
    \label{fig:materialFrameSch}
\end{figure}

\begin{example}[see \cref{fig:materialFrameSch}]
    \label{ex:materialFrame}
    Consider a $2$D periodic truss lattice generated by two primitive translation vectors encoded by a frame $\fref$~\parencite{HCBA2026}. In this architecture, the norms and directions of these vectors explicitly dictate the lengths, connectivity, and spatial positions of the beams composing the truss. Here, the reference frame $\fref$ at each node is not a measurement convention: it is a direct descriptor of the microstructure. Two distinct frames $\fref$ and $\frefp$ characterize two fundamentally different materials with distinct behaviors. If the constitutive behavior of such a truss includes non-linearities (in beams or joints), the energy density cannot be invariant under a change of reference frame $\fref$ because a single energy density cannot describe two distinct materials. In this context, $\fref$ is a parameter of the constitutive law, and variations of this frame letting the free energy density invariant are strictly restricted to the material symmetry group.
\end{example}

The ambiguity regarding the status of the reference frame $\fref$ (given by a the reference moving frame $\Sref$) as a measurement tool (\cref{ex:gaugeFrame}) or as a material parameter (\cref{ex:materialFrame}) explains why the arbitrary choice of a reference generalised configuration, is rarely discussed in standard mechanical and geometric literature. In the geometric theory of dislocations~\parencite{Kro1968,YG2012}, the moving frame is linked to the crystal lattice. The directors are modeled as "directors of matter" (e.g., lattice vectors defining slip planes). In such frameworks, the frame is endowed with a physical reality: it tracks the local, potentially defected, arrangement of the crystal lattice. Because the geometry is built to model the defects of a predetermined microstructural architecture, the freedom to choose an arbitrary reference frame at each point without altering the physics is not formalised. However for generalised or micromorphic media in finite strains, the identification of the moving frame $\Sref$ as a \emph{gauge choice} (as in \cref{ex:gaugeFrame}) must prevail: the framework, similar to classical continuum, should be ``blind'' to the micro-structure (see \cref{ex:cubeDef})~\parencite{Ger1973}.

\begin{example}[see \cref{fig:DiagGaugeInv}]
    \label{ex:cubeDef}
    Recalling \cref{ex:cubeConfig} about a small cube, in which its reference state is described by its center $\Xref$ and a trihedron drawn on it (a choice of frame $\fref$) representing the reference generalised configuration $\Coref$. After a uniform deformation of the cube, the new position of its center is $\Xdef$ and the new directions of the trihedron is a frame $\fdef$, representing the generalised deformed configuration $\Codef$. The mapping from $\Coref$ to $\Codef$ gives a $0$ dimensional generalised deformation $\Trans$. It is possible to conduct the \underline{same experiment}, \textit{i.e.} with the same $3$D deformation of the cube, but starting with a different reference frame $\frefp$ related to the first reference frame $\fref$ by a change of basis matrix $\matA\in\GL_3(\RR)$ ($\frefp=\fref\matA$). The new deformed frame $\fdefp$ is \underline{naturally} given by $\fdef\matA$ justifying that the generalised deformation $\Trans$ is a fiber bundle isomorphism (see \cref{rem:extensionMorphi}). From a mechanical point of view in which the free energy of the system is accounted at each macroscopic point (here $\Xref$), the deformed frames $\fdef$ and $\fdefp$ are given by two different observed generalised deformations $\Field(\Xref)=\Trans(\fref)$ and $\Fieldp(\Xref)=\Trans(\frefp)$. The relation between these two observed generalised deformations is given by (see the change of gauge formula \cref{eq:changeGaugeFormField}),
    \begin{equation*}
        \Fieldp(\Xref)
        =
        \Field(\Xref)\matA.
    \end{equation*}
    Since drawing a frame on the cube is only \underline{an observation}, the free energy stored by the cube during both experiments must be the same and therefore its expression in terms of the observed generalised deformation (and reference frame) should be the same when using $\Field$ or $\Fieldp$ with $\fref$ or $\frefp$ as reference frames.
\end{example}

Continuum mechanics is a system of beliefs~\parencite{KD2021}: the \underline{underlined} terms in \cref{ex:cubeDef}, which are true when considering a zero dimensional generalised deformation representing a $3$D homogenous deformation, are chosen to remain valid for a three dimensional generalised continua. If the micromorphic theory is to serve as a general continuum model capable of describing homogenised macroscopic behaviors independently of a specific microstructure, the reference moving frame $\Sref$ must be stripped of physical identification. It must be considered as a kinematic descriptor: a localised, user-defined choice of trihedron that is not defining the micro-structure as in \cref{ex:materialFrame}. In the formulation of micromorphic media using a trivialisation (\cref{eq:EnTotCTriCompa}), the reference moving frame $\Sref$ described by its matrix coordinates $\Gref$ must be considered as an arbitrary choice of \emph{gauge} used for the measurement of a kinematic variable: the micro-deformation (in its matrix form) given by,
\begin{equation}
    \label{eq:tframeInv}
    \TframeX
    =
    \Gdef(\Xdef)\Gref(\Xref)^{-1},
    \quad
    \Xdef = \trans(\Xref),
\end{equation}
in which $\Gdef$ is the matrix coordinates of the deformed moving frame $\Sdef$. Furthermore, it justifies the choice of the observed generalised deformation $\Field$ as the fundamental variable since the coordinates matrix of $\Field$ is given by $\Gdef\circ\trans$ (see \cref{rem:extensionMorphi}). It explains why the micro-deformation is considered as a fundamental variable of micromorphic media in the mechanical literature on generalised continua~\parencite{Eri1999}.

\begin{remark}
    Formulating an invariance principle with respect to a change of the reference generalised configuration $\Sref$ might, at first glance, raise concerns about its compatibility with material anisotropies. A parallel can be drawn with the principle of material frame-indifference, where an inappropriate formulation can mistakenly lead to enforcing isotropy~\parencite{Liu2005}. However, the material anisotropy observed macroscopically or microstructurally is not encoded by the choice of the gauge $\Sref$ (or its matrix representation $\Gref$), but rather by the layout of the microstructure relative to the fixed canonical reference frame $\Sglob$ (the observer). Once this global reference frame is established, anisotropy is fully captured through the specific functional dependence of the energy on the matrix representation of kinematic descriptors (the micro-deformation $\tframe$ and the gradient of the classical deformation $\gradT$). Since the proposed gauge transformation preserves the intrinsic geometric field $\tframe$, the underlying material symmetries remain strictly unaffected. This can be also seen in the fact that changing $\Sref$ does not affect $\gradT$ which is encoding the possible anisotropy of the media.
    %The alternative case where $\Sref$ itself would act as a physical variable, and where the energy would remain invariant under a specific subgroup of local transformations, belongs to the theory of generalised material symmetries, where a family of microstructures parameterised by these matter directors could share identical properties.
\end{remark}

% -------------------------------------
\subsection{Formulation of an invariance principle}
\label{sec:FormalPrinc}
% -------------------------------------

The requirement described in the previous sub-section (\cref{sec:DiscExample}) that any choice of reference moving frame, even depending on space, can be taken to measure the same relative change of frame. Besides, this change of frame should generalise to arbitrary reference moving frame $\Sref$. Recalling \cref{eq:mappingMoving} (and \cref{eq:mappingMoving}), it exists a unique (local) fiber bundle isomorphism $\DiffA$ (defined for all frames on $\coref$) mapping a reference moving frame $\Sref$ to an arbitrary moving frame $\Srefp$. This fiber bundle isomorphism, called a \emph{local gauge change}, thus verifies,
\begin{equation*}
    \DiffA(\Sref(\Xref))
    =
    \Srefp(\Xref),
    \quad \text{and} \quad
    \DiffA(\fref\mathbf{B})
    =
    \DiffA(\fref)\mathbf{B},
    \quad \forall\mathbf{B}\in\GL_3(\RR).
\end{equation*}
Local gauge changes are equivalently given by equivariant matrix valued functions $\matAL:\FEuc_{\vert\coref}\to\GL_3(\RR)$~\parencite[Thm 3.2.2]{Ble1981}, called \emph{local gauge transformations}, using
\begin{equation}
    \label{eq:defLocalTransfo}
    \DiffA(\fref)
    :=
    \fref\matAL(\fref)
    \quad \text{with for all $\mathbf{B}\in\GL_3(\RR)$, }
    \matAL(\fref\mathbf{B})
    =
    \mathbf{B}^{-1}\matAL(\fref)\mathbf{B}.
\end{equation}
The equivariance property (by the adjoint action of the structural group) of local gauge transformations means that knowing the value of $\matAL$ at only one frame $\fref$ with origin $\Xref$ determines its value for all frames at this point. In the following, we denote by $\matA:\coref\to\GL_3(\RR)$ the value of the gauge transformation $\matAL$ on the canonical frame $\Sglob$, \textit{i.e.} $\matA(\Xref):=\matAL(\Sglob(\Xref))$. Since the local gauge change $\DiffA$ is a fiber bundle isomorphism, it can be expressed in the trivialisation (\cref{eq:defTrivMorphism}) induced by the canonical frame $\Sglob$ by,
\begin{equation}
    \label{eq:defChangGauge}
    \begin{array}{rccc}
        \DiffA\colon
         & \coref\times\GL_3(\RR)
         & \to
         & \coref\times\GL_3(\RR)
        \\
         & \begin{pmatrix}
               \Xref \\ \gref
           \end{pmatrix}
         & \mapsto
         & \begin{pmatrix}
               \Xrefp := \Xref \\
               \grefp := \matA(\Xref)\gref
           \end{pmatrix}
    \end{array}
\end{equation}
in which $\matA:\coref\to\GL_3(\RR)$ can be thus interpreted as the matrix coordinates of the image of the canonical frame by the gauge change. Therefore, a ``general'' change of frame is given by the action of gauge changes on moving frames defined by,
\begin{equation}
    \label{eq:defactionChange}
    \DiffA\star\Sref
    :=
    \DiffA\circ\Sref.
\end{equation}
It naturally extends to generalised placements $\Pref:=\Sref\circ\pref$ and leaves the classical placement $\pref$ un-affected, \textit{i.e.}
\begin{equation}
    \label{eq:changeOfGen}
    \DiffA\star\Pref
    :=
    \DiffA\circ\Pref
    =
    \left( \DiffA\star\Sref \right)\circ\pref.
\end{equation}

\begin{remark}
    \label{rem:gaugeGroup}
    The set of all local gauge changes $\DiffA$ over $\coref$ forms a sub-group of the diffeomorphism of $\FEuc_{\vert\coref}$ (all frames on $\coref$, see \cref{rem:projection}) and is called the \emph{gauge group} $\GA{\coref}$ over $\coref$.
\end{remark}

\begin{figure}[h]
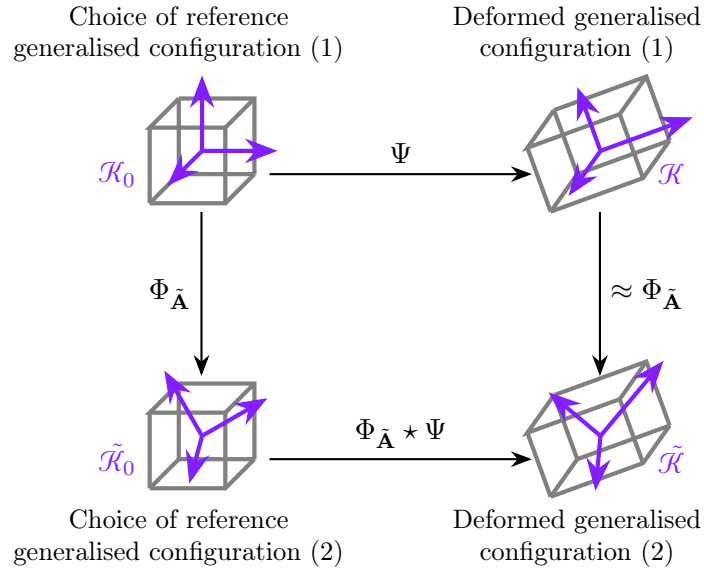

    \centering
    \tikzsetnextfilename{diagGaugeInv}
    \insertDiagGaugeInv
    \caption{Diagram showing the invariance by change of gauge discussed in \cref{sec:DiscExample}.}
    \label{fig:DiagGaugeInv}
\end{figure}

The action of \cref{eq:defactionChange} describes a change of frame, depending on space, on the reference configuration. Equivalently, \cref{eq:changeOfGen} describes a change of generalised reference configuration above the same classical configuration. In \cref{sec:DiscExample}, it has been establish that the free energy should be invariant if both the reference frame and the deformed frame are changed in the "same" way. Since the generalised deformation $\Trans$ is a fiber bundle isomorphism, looking at the image of a twisted reference frame $\fref\matA$ is equivalent to look at the image of the initial frame twisted by $\matA$, \textit{i.e.} $\Trans(\fref\matA)=\Trans(\fref)\matA$ as illustrated in \cref{fig:DiagMorphism}. Extending this idea to a local gauge change $\DiffA$ allows to define an action (push-forward by the inverse) of gauge changes on generalised deformations by,
\begin{equation}
    (\DiffA\star\Trans)(\fref)
    :=
    (\Trans\circ\DiffA)(\fref)
    =
    \Trans(\fref)\matAL(\fref),
    \quad
    \fref\in\FEuc_{\vert\coref}.
\end{equation}
It corresponds to the following \emph{action on the observed generalised deformation} $\Field:=\Trans\circ\Sref$ with $\Sref$ the reference associated moving frame (see \cref{fig:DiagGaugeInv} for illustration),
\begin{equation}
    \label{eq:actionFunc}
    \left( \DiffA\star\Field \right)(\Xref)
    :=
    \Field(\Xref)\matAL(\Sref(\Xref)),
    \quad \Xref\in\coref.
\end{equation}

\begin{remark}
    The action of local gauge changes $\DiffA\in\GA{\coref}$ on the observed generalised deformation in \cref{eq:actionFunc} can be expressed using a gauge change $\DiffAp$ defined on the deformed configuration, \textit{i.e.} $\DiffA\star\Field:=\DiffAp\circ\Field$ (explaining the $\approx\DiffA$ in \cref{fig:DiagGaugeInv}). The gauge transformation $\matALp:\FEuc_{\vert\codef}\to\GL_3(\RR)$ associated to $\DiffAp$ expresses using $\matAL:\FEuc_{\vert\coref}\to\GL_3(\RR)$, the gauge transformation associated to $\DiffA$, by
    \begin{equation*}
        \matALp(\Trans(\fref))
        :=
        \matAL(\fref),
        \quad
        \forall\fref\in\FEuc_{\vert\coref}.
    \end{equation*}
\end{remark}

The invariance described in \cref{sec:DiscExample} states that a choice of a generalised configuration above a given configuration is arbitrary: the free energy associated to an observed generalised deformation $\Field$ on a generalised deformation $\Coref:=\Pref(\BB)$ must be the same than the free energy associated to the observed generalised deformation $\DiffA\star\Field$ of \cref{eq:actionFunc} on the generalised reference configuration $\Corefp=\Prefp(\BB)$ of \cref{eq:changeOfGen} (illustrated in \cref{fig:DiagGaugeInv}). Since this change leaves the classical placement $\pref$ un-affected, this principle translates into the following invariance of the free energy of the system (from \cref{eq:EnTotField}) for any associated reference moving frame $\Sref$ on the reference classical configuration $\coref:=\pref(\BB)$ and any observed generalised deformation $\Field$ as,
\begin{equation}
    \label{eq:invGglob}
    \EnTot_{\pref,\DiffA\star\Sref}[\DiffA\star\Field]
    =
    \EnTot_{\pref,\Sref}[\Field],
    \quad
    \forall\DiffA\in\GA{\coref} \text{ (see \cref{rem:gaugeGroup})}.
\end{equation}
\begin{remark}
    \label{rem:gaugeinv}
    The invariance principle of \cref{eq:invGglob} will be referred as a \emph{gauge invariance principle} since it is a functional invariant by the action of the gauge group. Rigorously in gauge theories, gauge invariance is always formulated with a connection as a primary variable \parencite{Ble1981,Ham2017} which is not our point of view since we consider that the connection is fixed by the Euclidean space.
\end{remark}
This \emph{gauge invariance principle} (see \cref{rem:gaugeinv}) must be true for any classical reference placement $\pref$, thus it must be true for arbitrary small classical reference configuration $\coref$. Therefore, the gauge invariance of the free energy of the system (\cref{eq:invGglob}) is supposed to be true locally, \textit{i.e.} at every point $\Xref$ of a classical configuration $\coref$. It is the same argument than in classical continuum mechanics with material frame indifference (see \cref{sec:Recal}). It means that the free energy density $\EnLoc$ of \cref{eq:EnTotField}, verifies for all point $\Xref\in\coref$, any reference moving frame $\Sref$ around this point and any observed generalised deformation $\Field$ defined around this point, the following gauge invariance for all gauge changes $\DiffA$ around $\Xref$,
\begin{equation}
    \label{eq:invGloc}
    \EnLoc\left(
    \left(\DiffA\star\Sref\right)(\Xref),
    \left(\DiffA\star\Field\right)(\Xref),
    T_{\Xref}\left( \DiffA\star\Field \right)
    \right)
    =
    \EnLoc\left(
    \Sref(\Xref),
    \Field(\Xref),
    T_{\Xref} \Field
    \right).
\end{equation}

\begin{remark}
    This expression can be written compactly by forgetting some dependencies over $\Xref$,
    \begin{equation*}
        \EnLoc\left(
        \DiffA\star\Sref,
        \DiffA\star\Field,
        T\left( \DiffA\star\Field \right)
        \right)
        =
        \EnLoc\left(
        \Sref,
        \Field,
        T \Field
        \right).
    \end{equation*}
\end{remark}

The action on observed generalised deformation $\Field$ (\cref{eq:actionFunc}) reactivates the equivalence established in \cref{rem:fieldsEquiGauge} and, with it, resolves at once three difficulties left open earlier in this discussion. Since the generalised deformation $\Trans$ (equivalently $\tframe$) is unaffected by a change of gauge: only its observed generalised deformation $\Field$ in a moving frame $\Sref$, the representation through which the energy is actually evaluated, needs to transform. This is why the free energy of \cref{eq:EnTotField} is postulated on $\Field$ rather than on $\Trans$ itself, and it is what \cref{rem:extensionMorphi} anticipated: applying the ``same'' isomorphism to a twisted frame $\frefp:=\fref\matA$ yields a twisted image $\fdefp:=\fdef\matA$, and the same principle, once $\matA$ is let to vary over $\coref$ through a local change of gauge $\DiffA$, is what \cref{fig:DiagGaugeInv} now expresses as an assumption on the free energy.

The hypothesis that the theory must be blind to the choice of $\Sref$ (as demonstrated in \cref{ex:cubeDef}), is exactly what justify the extension of $\Trans$ to a fiber bundle isomorphism defined over all frames on the classical configuration, rather than at the single value $\Sref(\Xref)$ prescribed by the generalised reference configuration. The alternative, in which $\Sref$ (or equivalently $\Gref$) would be treated as a material variable like $\Xref$ (\cref{ex:materialFrame}), would instead confine the free energy density to be defined for that single value, turning $\Gref$ into a hidden parameter of the constitutive law, indexed by $\Xref$ like any other source of material inhomogeneity.

Thus, the extension of the generalised deformation $\Trans$ to a fiber bundle isomorphism, the postulate of the invariance of the free energy, and the postulated action on the observed generalised deformation $\Field$ are therefore three faces of the same hypothesis (illustrated in \cref{ex:cubeDef}) rather than three independent modelling choices.

% ------------------------------------------------------
\section{Consequences of the gauge invariance}%
\label{sec:Gauge}
% ------------------------------------------------------

Once a global canonical frame $\Sglob$ has been chosen, the free energy of the system and its density can be written in terms of the classical deformation $\trans$ and the micro-deformation $\tframe$ (see \cref{eq:EnTotCTri}). The trivialised free energy of the system $\EnTotC$ (and its density $\EnLocC$), depends on the classical placement $\pref$ and the associated reference moving frame $\Sref$ through its matrix coordinates $\Gref:\coref\to\GL_3(\RR)$. By denoting $\matA:\coref\to\GL_3(\RR)$ the matrix coordinate of the canonical frame by the change of gauge $\DiffA$, the matrix coordinates of the moving frame $\DiffA\star\Sref$ (see \cref{eq:defactionChange}) is given by (using \cref{eq:defChangGauge}),
\begin{equation}
    \label{eq:actGFrameTriv}
    (\matA\star\Gref)(\Xref)
    :=
    \matA(\Xref)\Gref(\Xref),
    \quad
    \Xref\in\coref.
\end{equation}

The gauge invariance have exactly been constructed such as the observed generalised deformations $\Field$ and $\DiffA\star\Field$ still correspond (under the action of gauge changes) to the same generalised deformation $\Trans$. Since the classical deformation $\trans$ and the micro-deformation $\tframe$ are quantities defined from $\Trans$ they are un-affected by the change of frames, it is the core concept of gauge theory (see \cref{rem:fieldsEquiGauge}). It means that the gauge invariance of \cref{eq:invGglob} can be written using the trivialised free energy of the system $\EnTotC$ in the following manner,
\begin{equation}
    \label{eq:invGglobC}
    \EnTotC_{\pref, \matA\star\Gref}[\trans,\tframe]
    =
    \EnTotC_{\pref, \Gref}[\trans,\tframe],
    \quad
    \forall\matA:\coref\to\GL_3(\RR),
\end{equation}
in which the matrix coordinates function $\matA\star\Gref$ is given by \cref{eq:actGFrameTriv}. Therefore, the local invariance of \cref{eq:invGloc} for a locally defined deformation $\trans$, a micro-deformation $\tframe$ and a matrix coordinates of a reference frame $\Gref$ around a point $\Xref$ expresses using the trivialised free energy $\EnLocC$ of \cref{eq:EnTotCTriCompa} as (using the notation introduced in \cref{eq:jetDefC}),
\begin{equation*}
    \EnLocC(j_{\Xref}(\trans,\tframe,\matA\star\Gref))
    =
    \EnLocC(j_{\Xref}(\trans,\tframe,\Gref)),
    \quad
    \forall\matA:\Xref\mapsto\matA(\Xref),
\end{equation*}
in which $\matA$ is the image of the canonical frame by locally defined gauge transformations $\matAL$ defined around $\Xref$. In this invariance, the deformation and the micro-deformation can be taken as fixed and the free energy only depends on the first gradient of $\Gref$. Therefore by fixing a deformation $\trans$ and a micro-deformation $\tframe$, the invariance can be written with a function $\funcR$ depending solely on the first gradient of $\Gref$ by,
\begin{equation*}
    \funcR(j_{\Xref}(\matA\star\Gref))
    =
    \funcR(j_{\Xref}(\Gref)),
    \quad
    \forall\matA:\Xref\mapsto\matA(\Xref),
\end{equation*}
in which $\funcR(j_{\Xref}(\Gref))=\EnLocC(j_{\Xref}(\trans,\tframe,\Gref))$. This writes equivalently for all matrix valued function $\matA$ defined around $\Xref$, as
\begin{equation}
    \label{eq:invLocSol}
    \funcR\left(
    \Xref, \matA(\Xref)\Gref(\Xref), \matA(\Xref)T_{\Xref}\Gref + \left(T_{\Xref}\matA\right)\Gref(\Xref)
    \right)
    =
    \funcR(\Xref, \Gref(\Xref), T_{\Xref}\Gref).
\end{equation}
Since this equation is true for any smooth function $\matA$ around $\Xref$ with values in $\GL_3(\RR)$, it is true for the following (affine) function defined for $\Xrefp$ a point in a neighbourhood in $\coref$ of $\Xref$,
\begin{equation}
    \matA(\Xrefp)
    =
    \mathbf{B} + \mathbb{B}\cdot\left( \Xrefp-\Xref \right),
\end{equation}
in which $\mathbf{B}=\Gref(\Xref)^{-1}$ and $\mathbb{B}=T_{\Xref}(\Gref(\Xref)^{-1})$ is  the differential of the inverse of $\Gref$ at $\Xref$. This is indeed a function from a neighbourhood of $\Xref$ with values in $\GL_3(\RR)$ because :
\begin{itemize}
    \item The determinant of $\matA(\Xrefp)$ is given by $\det(\mathbf{B})\det(\Id+\mathbf{u}(\Xrefp))$ in which $\mathbf{u}$ is a smooth function with values in $\mathrm{M}_3(\RR)$ (the Lie algebra of $\GL_3(\RR)$) such that $\mathbf{u}(\Xref)=\mathbf{0}$;
    \item Indeed, $\mathbf{u}$ is given by the left Maurer-Cartan form $\mathbf{B}^{-1}\delta\mathbf{B}$ in which $\delta\mathbf{B}=\mathbb{B}\cdot\left( \Xrefp-\Xref \right)$ is a tangent vector to $\mathbf{B}$ in $\GL_3(\RR)$;
    \item Since $\mathbf{B}$ and $\Id$ are invertible and the determinant is a smooth function, the function $\matA$ is invertible around $\Xref$.
\end{itemize}
Using that $\mathbb{B}$ can be written as~\parencite[Thm~2.4.4]{Car1977},
\begin{equation*}
    \mathbb{B}\cdot\delta\Xref
    =
    -\Gref(\Xref)^{-1}(T_{\Xref}\Gref\cdot\delta\Xref)\Gref(\Xref)^{-1},
\end{equation*}
the invariance of \cref{eq:invLocSol} with the particular gauge transformation described by $\matA$ (the image of the canonical frame) gives,
\begin{equation}
    \funcR(\Xref, \Id, \mathbf{0})
    =
    \funcR(\Xref, \Gref(\Xref), T_{\Xref}\Gref).
\end{equation}
This particular gauge transformation can always be constructed for any matrix coordinates function $\Gref$ and the invariance holds for all gauge transformations, thus the function $\funcR$ is only a function of $\Xref$. Consequently, the trivialised free energy density $\EnLocC$ does not depend either on $\Gref$ nor its differential and the trivialised free energy of the system can be written as,
\begin{equation}
    \label{eq:finalEnLoc}
    \EnTotC_{\pref}[\trans,\tframe]
    =
    \int_{\coref} \EnLocC\left(
    \Xref,\trans(\Xref),\TframeX,
    \gradT_{\Xref}, T_{\Xref}\Tframe
    \right)\masref(\Xref)\vol_{\metRef},
\end{equation}
which is exactly the form found in the mechanical literature of micromorphic media~\parencite{For2006}. Importantly, this is a sufficient and necessary condition for the free energy density to be gauge invariant. Mechanically, this result demonstrates that generalised media are not the study of the deformation of material directors: rather, they describe the deformation of arbitrary trihedra whose initial arrangement does not encode the microstructure.

\begin{remark}
    This result is not related to the dependency of the classical free energy to the linear tangent map $T\pref:T\BB\to T\coref$ of the classical reference placement when the free energy is associated to a placement~\parencite{DGK2026}. Indeed, if an energy is postulated to be associated to a placement (and not a deformation), when a change of variables is performed to recover an energy associated to a deformation, the dependency over $T\pref$ cannot be removed. Besides, we would have a similar dependency over the linear tangent map $T\Pref:T\BB\to T\Coref$ of the reference generalised placement if the free energy where postulated to be associated to a generalised placement.
\end{remark}

This gauge invariance have been unnoticed or unformalised in both the mechanical and geometric literature. In the mechanical literature, the question of the dependency on the reference frame (through $\Gref$) is masked by three modeling perspectives:
\begin{itemize}
    \item In modern computational micromorphic mechanics~\parencite{NGLM2015, ARK+2022}, the matrix representation of the micro-deformation $\Tframe$ ($\mathbf{P}$ in these works) is directly postulated a priori as the primary kinematic variable. By injecting $\Tframe$ directly into the power of internal forces or the energy density, the underlying concept of material directors is effectively "forgotten". The question of how $\Gref$ varies cannot arise because $\Gref$ is never explicitly manipulated.
    \item In historical formulations of micromorphic media~\parencite{Eri1999}, the reference moving frame is fixed from the outset of the problem. Because the reference frame is treated as an immutable background setting that is never varied or perturbed, its mathematical role as a potential variable is hidden. The energy density appears independent of $\Gref$ simply because the structural consequences of an observer's change of gauge are never investigated.
    \item In micro-macro homogenisation schemes~\parencite{CLH+2014,AMI2017,Com2023}, the reference moving frame $\Gref$ potentially containing data on the micro-structure (such as a primitive cell of the lattice give the beam lengths and directions, see \cref{ex:materialFrame}) is directly incorporated into the effective energy density, either implicitly through the integration process or explicitly as fixed material constants. Consequently, the dual nature of the frame, acting simultaneously as a kinematic descriptor and as microstructural descriptor cannot be distinguished. The frame is treated strictly as a material parameter of the constitutive law, preventing any analysis of its status as an independent geometric variable.
\end{itemize}

In the geometric theories of defects and dislocations~\parencite{Kro1968,YG2012,CCL2025}, formulations avoids the need for gauge invariance. These theories are designed to model the localised arrangement of the crystal lattice, the directors are strictly ``directors of matter'' (e.g., slip plane vectors). The frame possesses an absolute physical reality. Changing the reference frame at a point corresponds to changing the local arrangement of the crystal. Therefore, having arbitrary gauge choice would require additional physical arguments.

For the community applying differential geometry to generalised continua, the absence of this invariance principle stems not from an oversight but from a modelling choice made further upstream. In these geometric approaches~\parencite{Rub2000,EE2007,Eps2026}, placements are defined directly as isomorphisms between geometric fiber bundles (principal or affine bundles) rather than as mappings of a classical body to a target space. The fundamental kinematic variable is thus, from the outset, a bundle isomorphism: the generalised deformation $\Trans$ itself~\parencite{ED1998}, already possessing, the equivariant structure discussed in \cref{rem:fieldsEquiGauge}. The free energy density is accordingly built to depend on the first jet (gradient) of $\Trans$, measuring how every generalised configuration deforms, rather than restricting its scope to the single configuration observed during a given (thought) experiment. This is precisely the extension motivated in \cref{rem:extensionMorphi}: for this literature, it is not an extension, but the starting point of the formalism. Consequently, the question of whether the free energy may depends on the reference frame $\Sref$ never arises, since $\Sref$ is never considered as a possible argument of the energy in the first place and, with it, the arbitrariness of drawing a reference trihedron on a macro-point (\cref{ex:cubeDef}).

\begin{remark}
    \label{rem:repEpstein}
    This is explicit in the identification underlying \cref{rem:LinkEpstein}. Selecting a bijection between a moving frame on the body $\BB$ and a moving frame on a classical configuration corresponds to fixing a particular gauge $\Sref$ once and for all, \textit{i.e.} to adopt the material variable interpretation of $\Sref$ (\cref{ex:materialFrame}) rather than its gauge interpretation (\cref{ex:gaugeFrame}), in the sense made precise in \cref{sec:Gauge}. Besides, since constitutive laws in \parencite{EE2007}'s formalism are stated directly on the body $\BB$ rather than on a reference configuration, this choice is never exhibited explicitly as a dependency on $\Sref$: it is absorbed into the definition of the body's material structure.
\end{remark}

% ----------------------------------------------------------------
\section{Classification of generalised continua of first order}
\label{sec:Classif}
% ----------------------------------------------------------------

In practical applications as mentioned in the introduction, modelers employ reduction procedures to limit either the number of degrees of freedom or the number of constitutive parameters. These simplification pathways where roughly put into four categories (see \cref{sec:introduction}):
\begin{itemize}
    \item \emph{Microscopic Kinematic Reduction}: specific degrees of freedom of the micro-volumes are removed (\textit{e.g.} Cosserat);
    \item \emph{Coupled Kinematic Reduction}: certain microstructural degrees of freedom lose their independence and become entirely determined by the macroscopic deformation field (\textit{e.g.} strain-gradient);
    \item \emph{Energetic Reduction}: specific kinematic mechanisms or strain measures are postulated to yield no pointwise energetic cost (\textit{e.g.} relaxed micromorphic);
    \item \emph{Constitutive Modeling Reduction}: algebraic relationships are enforced between material parameters, reducing the number of independent parameters (\textit{e.g.} micro-foam).
\end{itemize}
Given only the final energy functional of a generalised media, it is generally impossible to discern a priori which of the four reduction pathways (or which combination of them) was utilised to derive the model (see \cref{ex:indeterminate}). Without prescribing a specific microstructure, the mathematical formulation cannot distinguish between a kinematic constrain and an energetic penalty.

\begin{example}
    \label{ex:indeterminate}
    Consider a Cosserat continuum whose energy density exhibits no dependence on microstructural kinematic descriptors related to possible micro-dilation or micro-shear deformation of the micro-volumes. This absence can be justified through two different physical arguments:
    \begin{itemize}
        \item The corresponding microscopic degree of freedom is physically blocked (the micro-structure is composed of rigid elements), meaning it does not exist within the configuration space.
        \item The micro-shear and micro-dilatation degrees of freedom vary freely, but the underlying micro-mechanics are such that its activation stores zero, or negligible, elastic energy.
    \end{itemize}
\end{example}

To resolve some ambiguity of the status of some models and establish a rigorous mathematical taxonomy, we restrict the focus of this section to formalising the \emph{Microscopic Kinematic Reduction} and the \emph{Coupled Kinematic Reduction} will be discussed in \cref{sec:constrain} (the other methods would require its own work). To do so, we distinguish three concepts:
\begin{itemize}
    \item The \emph{Generalised Configuration Space}, which collects all possible generalised configurations of the medium;
    \item The \emph{Kinematic Parametrisation}, which represents the chosen field variables (degrees of freedom) used to map this space;
    \item The \emph{State Transformation Parametrisation}, which represents how these degrees of freedom evolve under finite strains (i.e., the specific matrix form of the micro-deformation).
\end{itemize}
Thus, a kinematic reduction is a geometric reduction of the underlying configuration space. The kinematic fields used in mechanical literature are parameterisations of this underlying manifold. Consequently, restricting the allowed microscopic movements amounts to restricting the generalised configurations to a lower-dimensional sub-manifold of the standard frame bundle.

To formalize the \emph{Microscopic Kinematic Reduction} without ambiguity, the generalised configuration space must possess properties compatible with continuum mechanics. For this space to consistently model a generalised medium, any two local frames $\fref, \frefp$ attached to the same material point must be related by a unique change of basis matrix $\matA$ belonging to a specific matrix Lie group $\subG \subset \GL_3(\RR)$: any admissible micro-state must be reachable from any other admissible micro-state, and this transition must be realised in exactly one way. Besides, generalised deformations within this configuration space must remain structural bundle isomorphisms: it must be possible to transition between the spatial operations on frames and the algebraic operations on matrices without introducing coordinate dependencies. Mathematically, it describes a sub manifold of the frame bundle $\FEuc$ on which the Lie group $\subG$ acts freely and transitively.

Thus, a kinematic reduction corresponds to restricting the structural group $\GL_3(\RR)$ to a relevant matrix subgroup $\subG$, called a \emph{structural group reduction}~\parencite{Ble1981}. This defines a $\subG$-structure: a sub-bundle of frames whose transitions functions are with values in $\subG$. The permissible subgroups $\subG$ capable of generating such structures are not arbitrary, they are closed subgroups that can be defined as the stabiliser (symmetry) group of a tensor defined on the Euclidean space $\Euc$~\parencite{Kob1972}. Classifications of these subgroups exist in the mathematical literature, notably for subgroups of $\SL_3(\RR)$ that have been applied to classify material symmetries in finite strains~\parencite{MD2004}. Our contribution consists of exporting this classification to the kinematics of generalised media of order one as presented in \cref{tab:classifOrder} along with a description of the allowed deformations of the frames (\textit{i.e.} micro-volumes).

The reduction of the space of generalised configurations procedure operates as follows (see \cref{ex:ReducCos}): a matrix Lie subgroup $\subG \subset \GL_3(\RR)$ that represents the authorised micro-kinematic mechanisms is first selected and then the admissible generalised configurations are restricted to an orbit of frames related exclusively by the action of $\subG$ on an a selected frame. Since the choice of the reference frame is user-defined and carries no physical energy (see \cref{sec:Gauge}), one can systematically select the simplest available orbit: the one containing the canonical frame $\Sglob$. Finally, the matrix representation of the micro-deformation $\Tframe$ is automatically guaranteed to belong to the Lie group $\subG$.

\begin{example}[Cosserat media $\subG=\SO(3)$]
    \label{ex:ReducCos}
    To restrict the change of micro-volume states to pure rotations and forbid all micro-shears and micro-dilatations, the special orthogonal group is selected as the structural group, \textit{i.e.} $\subG = \SO(3)$. An orthonormal reference frame is chosen, thus the resulting configuration space consists entirely of orthonormal frames. If a non-orthonormal reference frame had been chosen, the action of $\SO(3)$ generates a class of non-orthonormal frames maintaining fixed relative angles and lengths. However, since the reference frame is arbitrary it is possible to select the canonical frame which is orthonormal. Besides, two orthogonal frames are always related by a unique rotation matrix meaning that the space of generalised configurations can be reduced to the orthonormal frame bundle. Consequently, the matrix representation of the micro-deformation $\Tframe$ loses its $9$ independent coefficients and is strictly constrained to be a rotation matrix with $3$ independent components (independently of the parametrisation chosen).
\end{example}

\Cref{tab:classifOrder} associates the most common generalised continua to a matrix sub-group corresponding to the reduction of the structural group along with the action on frames. A complete exploration of all possible $\subG$-structures~\parencite{MD2004} with their physical interpretation would be valuable. Even without this complete classification some mechanical theories can be excluded from this classification like the relaxed micromorphic model which is known to belong to the case of an energetically reduced model (\emph{Energetic reduction}). However, the micro-strain model~\parencite{FS2006} usually presented as a microscopic kinematic reduction (\emph{Microscopic kinematic reduction}) in which only a deformation, without rotation, of the micro-volumes are considered does not fit in this classification. Indeed, in this model the micro-deformation is a symmetric positive defined matrix (induced from a polar decomposition in which the rotation part is set to the identity) which does not form a matrix sub-group: even if the product of two symmetric positive defined matrix remains positive defined it can lose its symmetry property, \textit{e.g.}
\begin{equation*}
    \begin{pmatrix}
        2 & 0 \\ 0 & 1
    \end{pmatrix} \begin{pmatrix}
        1 & \gamma \\ \gamma & 1
    \end{pmatrix}
    =
    \begin{pmatrix}
        2 & 2\gamma \\ \gamma & 1
    \end{pmatrix},
    \quad \text{with } \gamma^2<1.
\end{equation*}
The example above shows that mechanically, a symmetric shear (same shear value $\gamma$ along $\bm{e}_X$ and $\bm{e}_Y$) in the plane followed by dilation in the $\bm{e}_X$ direction creates a global rotation of the sample. However, it is possible to construct a model (that does not seems to appear in the literature) only accounting for anisotropic dilatations of the micro-volumes without considering rotations by using the group $\text{Diag}_3(\RR^*_+)$ of diagonal matrices with strictly positive eigenvalues. Notice that unlike the micro-strain model, it cannot account for shearing of the micro-volumes.

\begin{table}[h!]
    \centering
    \begin{tblr}{
        colspec = {X[1,c,m]X[0.5,c,m]X[1,c,m]}
        }
        \toprule
        First order media                     & Group $\TframeX\in\subG$ & Action on the frames                 \\
        \midrule
        Micromorphic~\parencite{Eri1999}      & $\GL_3(\RR)$             & Rotation; Deformation                \\
        Incompressible Micromorphic           & $\SL_3(\RR)$             & Rotation; Incompressible Deformation \\
        Micro-stretch~\parencite{Eri1999}     & $\mathrm{CO}(3)$         & Rotation; Isotropic Dilatation       \\
        Cosserat~\parencite{CC1909}           & $\SO(3)$                 & Rotation                             \\
        Micro-dilatations~\parencite{Cow1984} & $\RR^*_+\Id$             & Isotropic Dilatation                 \\
        Cauchy                                & $\{\Id\}$                & Fixed                                \\
        \bottomrule
    \end{tblr}
    \caption{Classification of first order media based on structural group reduction of the frame bundle $\FEuc$ in which: $\SL_3(\RR)$ is the special linear group (invertible matrices with unitary determinant) and $\mathrm{CO}(3)$ is the conformal group (matrices of the form $\lambda\act$ with $\act\in\SO(3)$ a rotation and $\lambda\in\RR^*_+$ a positive isotropic dilation). This classification extends with the same groups to planar system when considering the $2$D Euclidean space (see \cref{rem:2DEuc}).}
    \label{tab:classifOrder}
\end{table}

% ----------------------------------------------------------------
\section{Convected frames and the lifting of the deformation}
% ----------------------------------------------------------------

This section addresses strain-gradient continua which constitute an alternative class of generalised media. It naturally integrates into the proposed geometric framework by adopting the existing formulation of convected frames \parencite{BR2017,FS2025}. Within this setting, the core concept lies in the lifting of the classical deformation into a generalised deformation that transports the convected frames. This lifting serves as a foundational tool that enables the introduction of several kinematic and constitutive notions developed further in this section. In particular, it introduces a natural decomposition of the micro-deformation in a \emph{compatible} and \emph{incompatible} part and it is the natural tool to formulate material frame indifference in our framework.

% ------------
\subsection{Kinematics of strain gradient continuum}
\label{sec:strain}
% ------------

\begin{figure}[h]
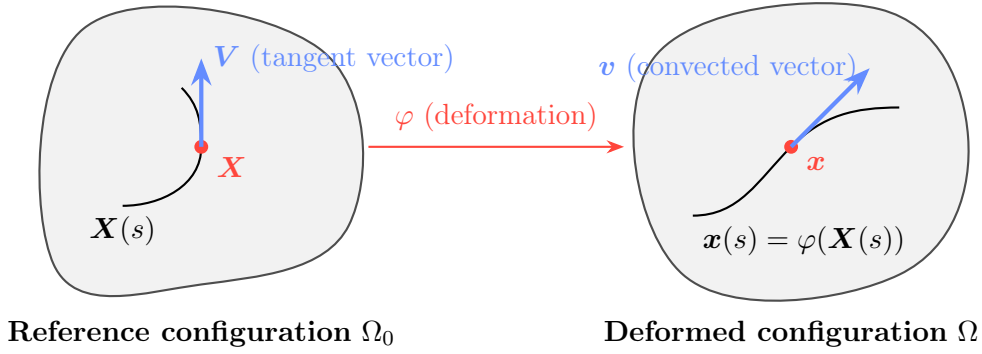

    \centering
    \insertDiagConvect
    \caption{Convected frames on a volume}
    \label{fig:convF}
\end{figure}

Let $\VV\in T_{\Xref}\coref$ be a tangent vector to a classical configuration $\coref$ at $\Xref$. Therefore it exists $s\mapsto\Xref(s)$ a smooth path of points in $\coref$ such as $\Xref(0)=\Xref$ and $\dot{\Xref}(0)=\VV$. Let $\trans:\coref\to\codef$ be a classical deformation mapping the reference configuration to the deformed configuration $\codef$. The \emph{convected} vector $\vv\in T_{\Xdef}\codef$ of $\VV$ which is a tangent vector to $\codef$ at $\Xdef:=\trans(\Xref)$ is naturally given by the gradient of the deformation $\gradT_{\Xref}$, \textit{i.e.} (see \autoref{fig:convF})
\begin{equation}
    \label{eq:convvec}
    \vv
    := \frac{\dd}{\dd s} \left( \trans(\Xref(s)) \right)_{s=0}
    = T_{\Xref}\trans\cdot\dot{\Xref}(0)
    = \gradT_{\Xref}\cdot\VV.
\end{equation}

For a $3$D media, a frame $\fref$ can be described by its basis vector $\VV_i=\fref(\bm{e}_i)$ (see \cref{eq:frame}), one can define the \emph{convected frame} $\fdef=\TransR(\fref)$ by the convected basis $\vv_i=\gradT_{\Xref}\cdot\VV_i$~\parencite{Ste2015,BR2017,FS2025}. The mapping of convected frame on a $3$D generalised configuration induces a particular generalised deformation $\TransR$ called the \emph{lifting} of the deformation $\trans$,
\begin{equation}
    \label{eq:lifting}
    \TransR(\fref)
    :=
    \gradT_{\Xref}\circ\fref,
    \quad
    \fref\in\pi^{-1}(\Xref), \Xref\in\coref \text{ (see \cref{rem:projection})}.
\end{equation}
Note that the micro-deformation associated to this generalised deformation $\TransR$ (\textit{i.e.} the tangent space isomorphism of \cref{rem:microDefTrans}) is exactly the gradient of the deformation $\gradT=T\trans$. It is the kinematic interpretation of a strain-gradient media~\parencite{FS2025} which recovers, when the free energy density of a micromorphic media (\cref{eq:finalEnLoc}) is selected with $\tframe=\gradT$, the usual energetic definition of strain-gradient media (a free energy density of second jet in $\trans$).

\begin{remark}
    The construction of convected frames is only possible for a 3-dimensional body: in \cref{eq:convvec} only tangent vectors to the reference configuration admits a path such that the vector is tangent to this path (see \cref{rem:Fmatrix}).
\end{remark}

In $3$D by \cref{eq:lifting}, the gradient of the deformation $\gradT$ and the micro-deformation $\tframe$ are of same nature (isomorphism of tangent space). However, only $\gradT$ is the tangent map of a diffeomorphism, we say that $\gradT$ is \emph{compatible}~\parencite{YG2012}.

% --------
\subsection{Decomposition of the micro-deformation}
\label{sec:decompMic}
% --------

For a generalised deformation $\Trans$ inducing a classical deformation $\trans$, it is possible to define a new generalised deformation $\TransId$ such that $\Trans = \TransR \circ \TransId$ in which $\TransR$ is the lifting of the induced classical deformation $\trans$ of \cref{eq:lifting}. This generalised deformation, that we decided to call the \emph{(Lagrangian) twisting deformation}, does not change the origins of the frames it acts on: both $\Trans$ and $\TransR$ induce the same classical deformation $\trans$.
\begin{equation*}
    \begin{mytikzcd}[ampersand replacement=\&]{diagLift}
        \FEuc_{\vert\coref} \arrow[r, "\TransId"] \arrow[dr, "\pi"] \arrow[rr, "\Trans", bend left] \&
        \FEuc_{\vert\coref} \arrow[r, "\TransR"] \arrow[d, "\pi"] \&
        \FEuc_{\vert\codef} \arrow[d, "\pi", swap]  \\
        \& \coref \arrow[r, "\trans"] \& \codef
    \end{mytikzcd}
\end{equation*}
The twisting deformation represents the generalised deformation applied to the frames of the generalised reference configuration such that the convected frames of those deformed frames give the generalised deformed configuration. Therefore in $3$D, a generalised deformation $\Trans$ can be uniquely expressed by the lifting of the deformation $\TransR:\FEuc_{\vert\coref}\to \FEuc_{\vert\codef}$ and the twisting deformation $\TransId:\FEuc_{\vert\coref}\to\FEuc_{\vert\coref}$ by,
\begin{equation}
    \label{def:twisting}
    \Trans(\fref)
    =
    \TransR(\TransId(\fref)),
    \quad
    \fref\in\FEuc_{\vert\coref}.
\end{equation}
\begin{remark}
    By uniquely, it means with a twisting deformation defined on the reference configuration. Indeed, we could have chosen a decomposition $\Trans=\TransId\circ\TransR$ in which the (Eulerian) twisting deformation would have been defined on the deformed configuration.
\end{remark}

We denote by $\tframeId$, that we decided to call the \emph{(Lagrangian) micro-twist}, the induced micro-deformation for the generalised twisting deformation $\TransId$ as defined in \cref{eq:defTangentIsoVec}, which is an isomorphism of the tangent space $T_{\Xref}\Euc=T_{\Xref}\coref$ at any point $\Xref$ of the classical reference configuration $\coref$,
\begin{equation}
    \label{eq:defmictwi}
    \tframeId_{\Xref}
    :=
    \TransId(\fref)\circ\fref^{-1},
    \quad
    \Xref=\pi(\fref).
\end{equation}
Note that this relation, also applied to the gradient of the deformation $\gradT$ and the micro-deformation $\tframe$, completely determines the generalised twisting deformation $\TransId$, \textit{i.e.} for any frame $\fref\in\FEuc_{\vert\coref}$ with $\Xref=\pi(\fref)$,
\begin{equation}
    \label{eq:tangentDecomp}
    \begin{cases}
        \Trans(\fref) = \tframe_{\Xref}\circ\fref, \\
        \TransR(\fref) = \gradT_{\Xref}\circ\fref, \\
        \TransId(\fref) = \tframeId_{\Xref}\circ\fref.
    \end{cases}
\end{equation}
The generalised deformation $\Trans$ decomposes into a \emph{convective part} $\TransR$ and a twisting part $\TransId$ (see \cref{def:twisting}) and using the induced tangent space isomorphisms defined in \cref{eq:tangentDecomp}. It induces a decomposition of the micro-deformation at any point $\Xref\in\coref$,
\begin{equation}
    \label{eq:decomMicro}
    \tframe_{\Xref}
    =
    \gradT_{\Xref}\tframeId_{\Xref}.
\end{equation}

\begin{remark}
    In the mechanical literature~\parencite{Eri1999,NF2007}, the inverse of the micro-twist is called the \emph{right stretch tensor} or the \emph{first Cosserat deformation tensor} since $\tframeId^{-1}=\tframe^{-1}\gradT$.
\end{remark}

%Naturally, a characterisation of a strain gradient kinematic is then given by an identity micro-twist, \textit{i.e.} $\tframeId_{\Xref}=\id_{T_{\Xref}\coref}$ (or $[\tframeId_{\Xref}]=\Id$). 
This decomposition has many analogies with the classical decomposition $\gradT^e\gradT^p$ used in finite strain metal plasticity~\parencite{Lee1969,Man1973}. The intermediate (or isoclinic) configuration used in plasticity can be interpreted in our formalism as a generalised configuration $\Codef_{\kappa}:=\TransId(\Coref)$, image of the reference generalised configuration by the twisting deformation. Besides, in the geometric theory of dislocations~\parencite{Kro1968,YG2012,CCL2025}, the plastic part $\gradT^p$ is \emph{incompatible} meaning that it is not the tangent map of a displacement field as to our micro-twist $\tframeId$. The incompatibility of $\gradT^p$ has been link to a continuous distribution of defects/dislocations within the material~\parencite{YG2012}.

% ----------------------------------------------------------------
\subsection{Material frame indifference}
\label{sec:MFI}
% ----------------------------------------------------------------

To formulate material frame indifference in statics on generalised continua, one must extends the action of isometries (diffeomorphism of the Euclidean space, see \cref{eq:defRigid}) to frames, \textit{i.e.} as diffeomorphisms of the frame bundle $\FEuc$. The straightforward way, is to consider for an isometry $\diffI$ its lifting $\Diff_{\diffI}$ given by $\Diff_{\diffI}(\fdef):=T_{\pi(\fdef)}\diffI\circ\fdef$ (see \cref{eq:lifting}). Once a canonical frame $\Sglob$ of the Euclidean space has been chosen and the isometry expresses as $\diffI(\Xdef)=\vtrans+\Rtrans\Xdef$ (with $\vtrans\in\RR^3$ and $\Rtrans\in\SO(3)$), its lifting $\Diff_{\diffI}$ expresses in the same trivialisation as,
\begin{equation}
    \label{eq:defRigFrame}
    \Diff_{\diffI}\colon
    \begin{pmatrix}
        \Xdef \\ \gdefp
    \end{pmatrix}
    \mapsto
    \begin{pmatrix}
        \Xdefp := \diffI(\Xdef) \\
        \gdefpp := \Rtrans\gdefp
    \end{pmatrix}.
\end{equation}
Mechanically, it means that isometries  rotates the frames describing the matter, \textit{i.e.} frames are convected along the isometries . Consequently, the material frame indifference of classical media of \cref{eq:invCla} extends to generalised continua (from \cref{eq:finalEnLoc}) by,
\begin{equation}
    \EnTot_{\pref}[\Diff_{\diffI}\circ\Field]
    =
    \EnTot_{\pref}[\Field],
    \quad
    \forall \diffI \text{ isometry of $\Euc$},
\end{equation}
for any classical placement $\pref:\BB\to\Euc$ inducing the classical configuration $\coref=\pref(\BB)$ and any observed generalised deformation $\Field:\coref\to\FEuc$. Using the trivialisation of the observed generalised deformation (\cref{eq:defTrivField}) involving the classical deformation $\trans$ and the matrix representation of the micro-deformation $\Tframe$ the new observed generalised deformation $\Diff_{\diffI}\circ\Field$ can be written as,
\begin{equation*}
    \Diff_{\diffI}\circ\Field\colon
    \begin{pmatrix}
        \Xref \\ \gref
    \end{pmatrix}
    \mapsto
    \begin{pmatrix}
        \Xdefp := (\diffI\circ\trans)(\Xref) \\
        \gdefpp := [T_{\trans(\Xref)}\diffI]\TframeX\gref
    \end{pmatrix},
\end{equation*}
in which $[T_{\trans(\Xref)}\diffI]$ is the matrix representation of the gradient of the isometry $\diffI$ at $\trans(\Xref)$ (given by $\Rtrans$ if $\diffI$ writes as in \cref{eq:defRigid}). Therefore, the material frame indifference writes in terms of the classical deformation $\trans$ and the micro-deformation $\tframe$ (see \cref{eq:finalEnLoc}) as,
\begin{equation}
    \label{eq:invMatFrame}
    \EnTot_{\pref}[\diffI\circ\trans, T\diffI\circ\tframe]
    =
    \EnTot_{\pref}[\trans,\tframe],
    \quad
    \forall \diffI \text{ isometry of $\Euc$}.
\end{equation}
Using the same argument than classical continuum mechanics (see \cref{sec:Recal}) that this invariance must be true for any reference placement meaning that it must be true locally. The invariance recasts as a pointwise invariance on the trivialised free energy density $\EnLocC$ of \cref{eq:finalEnLoc} by,
\begin{equation}
    \EnLocC(\Xref, \vtrans+\Rtrans\trans(\Xref), \Rtrans\TframeX, \Rtrans[\gradT_{\Xref}], \Rtrans T_{\Xref}\Tframe)
    =
    \EnLocC(\Xref, \trans(\Xref), \TframeX, [\gradT_{\Xref}], T_{\Xref}\Tframe).
\end{equation}
for all $\vtrans\in\RR^3$ and $\Rtrans\in\SO(3)$ given the expression of an isometry $\diffI$ in the trivialisation (see \cref{eq:defRigid}). This is the usual consequence of material frame indifference for micromorphic media used in the mechanical literature and we refer to \parencite{Eri1999,FS2006,NF2007} or \parencite{La2022} for some possible choices of a set of objective strain measures.

% ------------------
\section{Constrained media}
\label{sec:constrain}
% ------------------

In the mechanics of generalised continua, constrained media are traditionally introduced as higher-order models wherein the additional microstructural degrees of freedom lose their independence and are entirely determined by the macroscopic deformation field, \textit{i.e.} the \emph{coupled kinematic reduction} of \cref{sec:Classif}. Within the literature, these models are typically established by imposing an internal kinematic constraint linking the micro-deformation tensor $\tframe$ directly to the classical deformation gradient $\gradT$ usually in the form of an algebraic equation $\tframe=\mathcal{A}(\gradT)$~\parencite{Cap1985}. Consequently, constrained media operate as sub-models of second-grade theories. Despite the a priori freedom to choose an arbitrary mapping $\mathcal{A}$, mechanical and practical considerations have restricted the majority of literature to consider two cases: Strain-gradient media~\parencite{Min1964}, characterised by the identity mapping $\mathcal{A}(\gradT) = \gradT$, where the microstructure is perfectly convected along the macroscopic displacement and the Couple-Stress theory characterised by constraining the micro-rotation of a Cosserat media to match the macroscopic continuum rotation, a formulation pioneered by \parencite{Koi1963} and \parencite{MT1962} in infinitesimal strains (refered as the indeterminate couple-stress theory~\parencite{MGNM2016}).

From a theoretical perspective, it seems that a mathematical definition does not exist to derive an associated constrained media from any given higher-order theory. We propose a general method for first-order media based on the minimisation of the Euclidean distance (for matrices) between the matrix representations $\Tframe$ of the micro-deformation and the deformation gradient $\gradT$. Let $\subG \subset \GL_3(\RR)$ be a matrix subgroup corresponding to a reduced kinematics of the micromorphic model (see \cref{tab:classifOrder}). For a given matrix representation of a deformation gradient $[\gradT_{\Xref}]$, the associated constrained matrix representation of the micro-deformation $\TframeX_{\text{constr}}$ is defined as a projection of $[\gradT_{\Xref}]$ on $\subG$ that minimises the standard Euclidean metric for matrices (with Frobenius scalar product):
\begin{equation}
    \label{eq:MiniProb}
    \TframeX_{\text{constr}}
    \in
    \underset{\matA\in\subG}{\text{argmin}}
    \vert\vert \matA-[\gradT_{\Xref}] \vert\vert^2,
    \quad
    \vert\vert \matA \vert\vert^2 = \text{Tr}(\matA^T\matA).
\end{equation}
Mechanically, it corresponds to finding a frame deformation being the closest (in terms of matrix norms) to the induced macroscopic deformation (convected frames, see \cref{eq:convvec}) within the admissible generalised configurations. This projection method recovers the standard cases established in the literature for strain gradient media (trivially) and couple stress media. Indeed, when restricted to the orthogonal group ($\subG = \SO(3)$), the projection operation naturally coincides with the rotation part of the polar decomposition of $\gradT$ (see \cref{app:miniProb}).

\begin{remark}
    Mathematically, the minimisation problem of \cref{eq:MiniProb} is the problem of approximating a matrix of $\GL_3(\RR)$ by an element of a subgroup $\subG$ (minimisation on a Lie group). An alternative formulation of the problem involving the micro-twist $\tframeId$ (\cref{eq:defmictwi}) is,
    \begin{equation*}
        [\tframeId_{\Xref}]_{\text{constr}}
        \in
        \underset{\matA\in[\gradT_{\Xref}]^{-1}\subG}{\text{argmin}}
        \vert\vert \matA-\Id \vert\vert^2,
        \quad
        \text{with } [\gradT_{\Xref}]^{-1}\subG:=\left\{ [\gradT_{\Xref}]^{-1}\matA\,\vert\,\matA\in\subG \right\}.
    \end{equation*}
    This problem has a much simpler cost function however its minimization set has no longer a group structure: it is the orbit of $[\gradT_{\Xref}]^{-1}$ by the (right) action of the subgroup $\subG$. However, this formulation suggest that constrained media can be defined with constrains on the micro-twist rather than on the micro-deformation.
\end{remark}

Applying the minimisation problem of \cref{eq:MiniProb} to a micro-dilatation medium, where the admissible micro-deformations are restricted to pure dilatations ($\subG = \RR^*_+\Id$) with a positive determinant gradient of deformation $\gradT$ (corresponding to a deformation implying no volume flipping), leads to (see \cref{app:miniProb}),
\begin{equation*}
    \TframeX_{\text{constr}}
    =
    \frac{1}{3}\mathrm{Tr}([\gradT_{\Xref}]) \Id.
\end{equation*}
It corresponds to the hydrostatic component of the macroscopic deformation gradient $[\gradT_{\Xref}]$. An alternative pathway toward identifying such constrained sub-models is found in invariant theory and harmonic decomposition notably for micro-dilatation and micro-stretch media~\parencite{Auf2013}. However, these results are derived within the framework of infinitesimal strains using its harmonic decomposition. Remarkably, our minimisation procedure in finite strains (\cref{eq:MiniProb}) matches the small-strain result of \parencite{Auf2013} in which the \emph{dilataion-gradient} tensor is defined as the hydrostatic part of the micro-deformation. The result of this minimisation strategy on Cosserat and micro-dilatation media suggests that a systematic investigation of the optimisation problem over the entire classification of matrix Lie subgroups (at least those among \cref{tab:classifOrder}) and determining whether every valid microstructural $\subG$-structure leads to a unique kinematic constrain $\TframeX = \mathcal{A}([\gradT_{\Xref}])$ would provide an exhaustive classification. Besides, even if such a constrain exist, there is now garantee that the associated mechanical theory would be well-posed and the obtention of the equilibrium equation can only be done using a variational principle~\parencite{MGNM2016,FS2020}.

% ---------------
\section{Conclusion}
% ---------------

This work proposed a gauge-theoretic reformulation of generalised continuum mechanics, addressing in a unified manner two difficulties identified in the introduction:
\begin{itemize}
    \item The ambiguous physical status of the \emph{directors of matter}: whether they should be regarded as material parameters explicitly encoding a specific microstructural architecture, or regarded as arbitrary kinematic descriptors (\emph{gauges}) tracking the deformation of the meso-structure, independently of any particular underlying microstructure;
    \item The absence of a unified classification of generalised contina: the various simplification strategies used to render such theories operational (reductions or constrains of kinematic degrees of freedom, removing energetic contributions or adding relations between constitutive parameters) are only unambiguously distinguished when derived from a known microstructure.
\end{itemize}
Defining generalised configurations as moving frames (of the Euclidean space, \textit{i.e.} trihedra, independently of the dimension of the body) over classical configurations allowed us to establish that the free energy density used in micromorphic elasticity is the sufficent and necessary condition for the free energy associated to a change of generalised configuration to be independent with respect to the choice of a reference generalised configuration (\emph{gauge invariance}). It sets the status of the directors of matter in elasticity as arbitrary kinematic gauges rather than material descriptors tied to a specific microstructure, and simultaneously recovers the micromorphic model as the general-purpose theory built upon this principle~\parencite{Ger1973}. Furthermore, the structural group reduction of the configuration space provided a systematic taxonomy of first-order generalised media through the theory of $\subG$-structures, while strain-gradient continua were shown to emerge naturally via convected frames~\parencite{FS2025}. Finally, a coherent treatment of material frame-indifference and a procedure connecting first-order and second-grade models (constrained media) were proposed.

The present approach should not be understood as competing with generalised continua already available in the mechanical literature, but rather as an attempt to rationalise their theoretical structure by clarifying the geometric status of the kinematic variables at play and organising the resulting sub-theories within a single classification. Rather than introducing a new model among many, the framework developed here offers a common ground onto which the existing, and currently scattered, families of higher-order and higher-grade theories can be mapped and compared. In doing so, it also clarifies the nature and objectives of elastic generalised continuum theories: a general-purpose elastic theory is a theory of \emph{arbitrary descriptors} of microstructures, whose physical content is recovered only once a specific choice of a microstructural interpretation is made. This framework is also compatible with thin structure modelling since it is valid for body of dimension lower than $3$, an aspect only recently overlooked in a geometric manner~\parencite{JE2026}.

Several limitations of the present work open natural directions for future research. The classification proposed here is restricted to first-order generalised media, while its extension to higher-order frames~\parencite{FS2020}, inducing non linear actions of the structural group, appears conceptually possible, it is not pursued here, and is premature given that even first-order models are only beginning to find applicative use. The present study is also confined to quasi-statics: extending it to dynamics would proceed from a generalised placement path (as in classical continuum mechanics), but requires confronting the modelling of micro-inertia, a notoriously delicate question~\parencite{MNA+2017}. Finally, the formulation of consistent boundary conditions, a long-standing difficulty for generalised continua, whether addressed theoretically~\parencite{MGNM2016,IEFS2022}, numerically~\parencite{FCB2011}, or experimentally~\parencite{VRW+2026}, has been left entirely outside the scope of this paper and warrants a dedicated treatment.

More broadly, this work can be read in light of the model crisis described in~\parencite{IM2024}: the emergence, in architected materials, of behaviours that classical Cauchy continua cannot describe is an anomaly calling for a paradigm shift. It also concerns the tools used, favouring the principle of virtual power over historical routes such as conservation laws~\parencite{IS2020,IS2023} or Cauchy cuts~\parencite{dSM2012}. Internal energy formulations, as adopted here, are naturally compatible with this principle, which in turn accommodates non-integrable boundary conditions~\parencite{KD2021}. It is within this variational framework, enriched by the tools of gauge theory, that the present work is positioned.

% ----------------------------------------------------------------
\appendix
% ----------------------------------------------------------------

% --------------
\section{Some results on minimisation problem for constained media}
\label{app:miniProb}
% --------------

The goal of this section is to establish some solutions of the problem of \cref{eq:MiniProb}, \textit{i.e.}
\begin{equation*}
    \TframeX_{\text{constr}}
    \in
    \underset{\matA\in\subG}{\text{argmin}}
    \vert\vert \matA-[\gradT_{\Xref}] \vert\vert^2,
    \quad
    \vert\vert \matA \vert\vert^2 = \text{Tr}(\matA^T\matA),
\end{equation*}
in which $\subG$ is a matrix Lie subgroup of $\GL_3(\RR)$ and $[\gradT_{\Xref}]$ belongs to $\GL_3(\RR)^+$, the set of postive defined $3$ by $3$ invertible matrices. Indeed, for a meaningful deformation (no negative volume) the gradient of the deformation in a fixed canonical basis is necessarily positive defined. To shorten the notations, let us consider the following identical problem with $\mathbf{B}\in\GL_3(\RR)^+$,
\begin{equation}
    \matA^*
    \in
    \underset{\matA\in\subG}{\text{argmin}}
    \vert\vert \matA-\mathbf{B} \vert\vert^2,
    \quad
    \vert\vert \matA \vert\vert^2 = \text{Tr}(\matA^T\matA).
\end{equation}

\paragraph{Case $\subG=\SO(3)$}

The cost function $f(\matA)=\vert\vert \matA-\mathbf{B} \vert\vert^2$ can be written as (using that $\matA$ is a rotation),
\begin{equation*}
    f(\matA)
    =
    3-2 \mathbf{B}:\matA + \vert\vert \mathbf{B} \vert\vert^2.
\end{equation*}
in which $:$ is the scalar product for matrices, \textit{i.e.} $\matA:\mathbf{B}=A_{ij}B_{ij}$. Since $3+\vert\vert \mathbf{B} \vert\vert^2$ is strictly positive, the minimum of $f$ is archieved when the scalar product $\matA:\mathbf{B}$ is maximum. Combining the polar decomposition $\mathbf{B}=\act\mathbf{U}$, in which $\mathbf{U}$ is symmetric positive defined (unique since $\mathbf{B}$ is positive defined), and the diagonalisation of $\mathbf{U}=\mathbf{P}\mathbf{D}\mathbf{P}^T$ (Spectral theorem) leads to search for $\tilde{\matA}\in\SO(3)$ the maximum of,
\begin{equation}
    \label{eq:probPS}
    \mathbf{D}:\tilde{\matA},
    \quad
    \text{with }
    \tilde{\matA} = \mathbf{P}^T\act^T\matA\mathbf{P}.
\end{equation}
Since $\mathbf{U}$ is positive defined, all the terms of the diagonal matrix $\mathbf{D}$ are positive. Besides $\tilde{\matA}$ being a rotation, all its components are lower or equal to one. Consequently, $\mathbf{D}:\tilde{\matA}$ is maximal when all the diagonal terms of $\tilde{\matA}$ are equal to one which is only possible when $\tilde{\matA}=\Id$. Using the expression of $\tilde{\matA}$ in \cref{eq:probPS}, it leads to $\matA=\act$ the rotation part of the $\act\mathbf{U}$ decomposition of $\mathbf{B}$.

\begin{remark}
    This result is obtained as a particular case of the orthogonal Procuste problem~\parencite{Sch1966}.
\end{remark}

\paragraph{Case $\subG=\RR^*_+\Id$}

The cost function $f(\matA)=\vert\vert \matA-\mathbf{B} \vert\vert^2$ can be written as a scalar function $h$ on $\RR_+^*$ since $\matA\in\RR^*_+\Id$ is equivalent to $\matA=\lambda\Id$ with $\lambda\in\RR^*_+$, thus
\begin{equation*}
    h(\lambda)
    :=
    f(\lambda\Id)
    =
    3\lambda^2 - 2\lambda\text{Tr}(\mathbf{B}) + \vert\vert \mathbf{B} \vert\vert^2.
\end{equation*}
By differentiating this expression, $h$ admits a unique minimum (since $\ddot{h}(\lambda)=6>0$) at $\lambda=\frac{1}{3}\text{Tr}(\mathbf{B})$ which is always strictly positive since $\mathbf{B}$ is positive defined.

% ----------------------------------------------------------------
\printbibliography%[heading=bibintoc]
\end{document}